\documentclass{aa}
\usepackage{natbib}
\usepackage{xspace}
\usepackage{amssymb}
\usepackage{amsmath}
\usepackage{graphicx}
\usepackage{rotating}
\usepackage{natbib}

\begin{document}  
  
  \title{Coagulation, fragmentation and radial motion of solid particles in
    protoplanetary disks}

  \titlerunning{Coagulation and fragmentation of grains}
  
  \authorrunning{Brauer,
    Dullemond \and 
    Henning}
  
  \author{F.~Brauer \and 
    C.P.~Dullemond \and 
    Th.~Henning}
  
  \date{\today} 

  \institute{Max-Planck-Institut f\"ur Astronomie, K\"onigstuhl 17, D--69117
  Heidelberg, Germany; e--mail: brauer@mpia.de}
  
  \abstract{ The growth of solid particles towards meter sizes in
    protoplanetary disks has to circumvent at least two hurdles,
    namely the rapid loss of material due to radial drift and particle
    fragmentation due to destructive collisions.  In this paper, we
    present the results of numerical simulations with more and more
    realistic physics involved. Step by step, we include various
    effects, such as particle growth, radial/vertical particle motion
    and dust particle fragmentation in our simulations. We demonstrate
    that the initial dust-to-gas ratio is essential for the particles
    to overcome the radial drift barrier. If this value is increased
    by a factor of 2 compared with the canonical value for the
    interstellar medium, km-sized bodies can form in the inner disk
    ($<2$~AU) within $10^4$~yrs. However, we find that solid particles
    get destroyed through collisional fragmentation. Only with the
    unrealistically high-threshold velocities needed for fragmentation
    to occur ($>30$~m/s), particles are able to grow to larger sizes
    in disks with low $\alpha$ values. We also find that less than 5\%
    of the small dust grains remain in the disk after 1~Myrs due to
    radial drift, no matter whether fragmentation is included in the
    simulations or not. In this paper, we also present considerable
    improvements to existing algorithms for dust-particle coagulation,
    which speed up the coagulation scheme by a factor of $\sim 10^4$.}

\maketitle

\begin{keywords}
accretion, accretion disks -- circumstellar matter 
-- stars: formation, pre-main-sequence -- infrared: stars 
\end{keywords}

\section{Introduction}

The coagulation of sub-$\mu$m dust particles is believed to be the
initial step of planetesimal formation in disks around
pre-main-sequence stars \citep{hbook,natta07}. Evidence for dust grain
evolution beyond sizes that are found in the interstellar medium is
provided by mid-infrared spectroscopy of disks around Herbig Ae/Be
stars \citep{bouwman01,boekelaebe03}, T~Tauri stars
\citep{prz03,kessler07} and also around brown dwarfs
\citep{apai04,apai05,apai2007,sicilia07}. Millimeter interferometry
indicates large populations of particles which have grown to even
larger sizes, ranging up to several centimeters
\citep{tes03,Wil05,Rod06}.

All these observations give reason to model the evolution of particles
in protostellar disks in order to explain the observational data
\citep{DD04,DD05,tan05,nomura06,allesio06,ormel07}. However, these
theoretical investigations do not only attempt to model the evolution
of the appearance of protostellar disks. They also unveil certain
obstacles in the formation of planetesimals by particle coagulation
\citep{you04,domppv07,brauer07}.

One of these obstacles is the radial inward drift of solid bodies
towards the central star as first dicussed by \cite{Whipple72} and
\cite{Wei77}. The gas in a protostellar disk moves slightly
sub-Keplerian due to a radial pressure gradient. For this reason, the
dust which moves with near Keplerian velocity feels a continuous head
wind of gas. Hence, the particle loses angular momentum due to drag
forces between gas and dust and spirals inward.  If this radial drift
of the particles is not prevented by some mechanism, then the solid
particles drift into the inner evaporation zone and are lost for the
process of planetesimal formation. To give an example, the radial
drift time scale for meter-sized bodies at 1 AU is $\sim 10^2$
yrs. Within this time scale these boulders at 1 AU drift into the
inner regions of the disk and evaporate. A possible way out of this
problem is particle growth since the radial drift velocity is fairly
dependent on the particle radius. For example, the drift velocity of
meter-sized particles at 1 AU in the disk is $\sim 50$~m/s, but the
radial drift velocity of 10~m sized bodies is already 10 times
lower. Therefore, swift particle growth could prevent the particles
from drifting into the evaporation zone. However, the general disk
evolution comprises a considerable particle loss due to evaporation
which is hard to prevent. This problem is a major topic of this
paper. Other sublimation zones of the disk, e.g. the snow line at
$\sim 2$~AU \citep{lecar06}, could also play a role for particle drift
and coagulation processes. However, we will for now neglect this issue
which will be the topic of a forthcoming paper.

Another obstacle is the fragmentation of solid particles. While
low-velocity collisions lead to particle growth, high velocity impacts
lead to destruction \citep{bw98,poppe99,BluWur99}. For example, the
relative particle velocity of meter-sized bodies in a protostellar
disk can be more than $\sim 30$~m/s
\citep{Wei77,marmizvol91}. \cite{Benz00} found that meter-sized rocks
appear unlikely to survive an impact with a relative low collision
velocity of some cm/s. For porous objects, collision velocities higher
than 4\% of the sound speed lead to particle destruction
\citep{sir04}. For this reason, the particle size of roughly a meter
seems to pose an upper limit for particle coagulation.

These two obstacles, the radial drift barrier and the fragmentation barrier,
are the issue of this paper. We present a disk model including the growth, the
radial drift and the fragmentation of the particles. We show how these three
effects change the evolution of the disk by including them step by step.
\begin{enumerate}
\item In the first step we only consider particle coagulation due to Brownian
  motion, vertical settling and turbulent mixing. This step shows to which
  sizes particles can grow if radial drift and fragmentation are neglected.
\item The second step includes the radial drift and the radial mixing of
  dust. The particles are now allowed to move inwards and to disappear into
  the evaporation zone. However, we investigate which disk parameters
  influence the drift time scales and for which parameters the dust particles
  overcome the drift barrier.
\item The last step also includes particle fragmentation. We show under which
  conditions, i.e. in which regions of the disk and for which disk parameters,
  it is possible for the dust to overcome this barrier.
\end{enumerate}
The radial drift barrier is not only of interest for the radial drift
itself. Particles close to this barrier are most susceptible to the
motions of the gas and the gravitational effects of the dust. For
example, particles can be trapped in very elongated gas vortices in
magnetorotational turbulence \citep{balhaw91,bargesomm95}.  These
effects can slow down the radial drift by a factor of two
\citep{JohKlaHen06}. Under certain conditions the solid particle layer
itself may become gravitationally unstable \citep{anders07}. In high
dust density regions, the particles contract due to their own gravity
and may form a planetesimal within a few orbits
\citep{andersnat07}. Moreover, the flow of the gas and the dust can be
unstable to the streaming instablity \citep{yougoo05} which leads to
particle clumping, and possibly also to a gravitational collapse of
the dust. Apparently, the radial drift barrier is not only connected
to the radial motion of the dust particles, but involves various other
important effects as well. For this reason, it is vital to answer the
question if particles can actually reach the size regime at which
non-linear effects become of importance.

In this paper we will implement a 2+1 dimensional model. The first
dimension is the radial coordinate of the disk $r$, the second one is
the height above the midplane $z$ and the third coordinate is the mass
of the dust particles $m$. The dust may move radially due to radial
drift and radial mixing. We will numerically solve the continuity
equation for this problem for each particle species. The time
evolution for the particle size distribution is determined by the
coagulation equation. We will numerically solve this equation as
well. In the vertical direction, we will always assume that each
particle species is in vertical sedimentation/mixing equilibrium
\citep{dms95,food}. Hence, we will not solve the time dependent
continuity equation in the vertical direction as done for example by
\cite{DD05}. Nor do we need to solve the coagulation equation at all
$z$ explicitly. Instead, we solve the vertically integrated
coagulation equation, which significantly saves computational time
(cf. Appendix \ref{vertint}). We also formulated the coagulation
equation in an implicit way (cf. Appendix \ref{idiff}) which saves
another factor of $\sim 100$ of computer simulation time.

\section{Model equations}\label{moddel}
\subsection{Disk model}

We consider a disk of mass $M_{\mathrm{disk}}$ and an inner and an outer disk
radius of $r_{\mathrm{in}}=0.03$~AU and $r_{\mathrm{out}}=150$~AU. We adopt a
central stellar mass of $0.5$ $M_{\odot}$ and a disk mass of 0.01 $M_{\star}$
if not otherwise noted. The mass distribution of gas and dust inside the disk
is given by the surface density $\Sigma$. This quantity generally depends on
the distance to the central star $r$ and the azimuthal angle
$\varphi$. However, since we assume a disk which is axisymmetric the surface
density $\Sigma$ only depends on the radius $r$. We assume that this
dependency can be described by a power law
\begin{equation}
\Sigma=\Sigma_0\left(\frac{r}{1 \mathrm{AU}}\right)^{-\delta}\;.
\end{equation}
The power law index of the surface density $\delta$ is set to be 0.8
in the course of this paper following \cite{Kit02} and
\cite{andwill07}. The surface density at 1~AU, which is denoted by
$\Sigma_0$, is chosen in a way that the condition
\begin{equation}
2\pi\int_{r_{\mathrm{in}}}^{r_{\mathrm{out}}}\Sigma(r)rdr=M_{\mathrm{disk}}
\end{equation}
holds. With the values mentioned above, the surface density of the gas
at 1~AU in the disk can be calculated to be $\sim 20$~g/cm$^2$. 

\begin{figure}
\begin{center}
\includegraphics[scale=0.5]{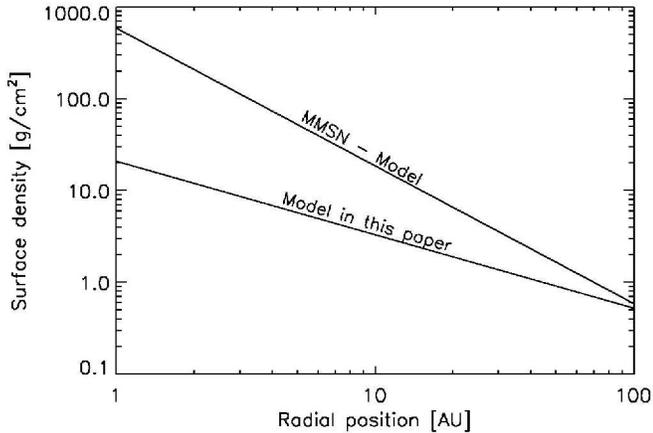}
\caption{The surface density distribution of the gas as a function of
  disk radius for the disk model discussed in the paper at hand and
  the MMSN model as discussed in Sec.~\ref{moddel}. Note, that the
  surface densities at 1~AU differ by more than one order of
  magnitude.\label{hmod}}
\end{center}
\end{figure}

This surface density distribution of gas and dust, on which the dust
particle coagulation calculations in this paper are based, differs
significantly from the disk model which is usually referred to as the
minimum mass solar nebula (MMSN) model \citep{wei77hay,hayashi81}. The
fundamental difference can be found in the distribution of the mass of
gas and dust in the disk. In the MMSN disk model, the power law index
of the surface density is $\delta=1.5$, while the disk model in this
paper adopts $\delta=0.8$. Fig.~\ref{hmod} shows the surface densities
as a function of disk location for both disk models assuming a disk
mass of $0.01$~$M_{\star}$.
The MMSN model implies surface densities of $\sim 600$~g/cm$^2$ at
1~AU in the disk. With the disk model in this paper, we yields surface
density values of $\sim 20$~g/cm$^2$ which is more than one order of
magnitude lower. 

The actual distribution of mass in a protoplanetary nebula is still a
matter of debate. There is evidence from meteoritics that the
densities in the protosolar nebula in the planet forming region have
been very high, implying disk masses much larger than the MMSN
\citep{desch02}. On the other hand, resolved millimeter dust emission
maps of protoplanetary nebula seem to indicate much lower surface
densities \citep{andwill07}. However, millimeter dust observations of
disks may not trace the radial profile of the gas mass density
correctly since particle growth to larger sizes is expected to proceed
more quickly in the inner parts of the disk than in the outer
parts. For this reason, the dust continuum emission becomes flat even
though the radial profile of the surface density might have a steep
radial behaviour. Hence, analysing dust emission maps assuming a
constant dust particle size throughout the disk likely leads to power
law indices $\delta$ which are systematically shifted towards lower
values.

The actual surface density distribution probably lies in between these
two extreme cases, i.e. the MMSN model with $\delta=1.5$ and the
observational median of $\delta=0.5$ \citep{andwill07}. The surface
density profile adopted in this paper is chosen to be between these
two extremes. Our goal is to gain insight in dust disk evolution in a
environment which might well be a likely scenario considering the
$\delta$ range under discussion.

We assume that the gas in the disk is in a steady state, even for
times as long as 1~Myr. Hence, the gas densities in our model do not
change in time. We only focus on the dust component in the disk which
evolves on a steady gas background. To unveil the robustness of this
assumption, we compare the following time scales. Particle growth time
scales are of the order of $\sim 10^2$~orbital times before
fragmentation prevents further particle growth
(c.f. Sec.~\ref{crf}). Radial gas accretion velocities are of the
order of $\sim 1\ldots 10$~cm/s at 1~AU in the disk
\citep{TakLin02}. With regard to 1~AU, this leads to accretion time
scales of the order of $10^5$~yrs, which is much larger than typical
particle growth time scales. For example, \cite{TakLin02} find that in
the first $10^{4\ldots 5}$~yrs, the gas surface density between 1 and
100~AU is hardly affected by viscous evolution. However, after
$~10^5$~yrs, the surface density of the gas may change significantly
over time scales of several Myrs \citep{reyes07}. This introduces a
systematic uncertainty in our dust evolution model regarding late
evolutionary stages of T~Tauri disks.

The temperature T is assumed to be the midplane temperature of a disk
irradiated under an angle of $\alpha_{\mathrm{irr}}=0.05$ around a
T~Tauri star with a surface temperature of $T_{\star}=4000$ K and a
radius of $R_{\star}=2.5$ $R_{\sun}$. If we assume the disk to be
isothermal in the vertical direction then the temperature is given by
\begin{equation}\label{temp}
T=\alpha_{\mathrm{irr}}^{1/4}\left(\frac{r}{R_{\star}}\right)^{-1/2}
T_{\star}\;.
\end{equation}
With this dependency the temperature at 1 AU is 204~K. The evaporation
temperature at 0.03 AU is $\sim 1400$~K. 

\subsection{Vertical structure of gas}

We consider a thin disk, which means that $z\ll r$. The quantity $z$ denotes
the height above the midplane. Under this condition the vertical mass density
distribution of the gas can be described by
\begin{equation}\label{topf}
\rho_{\mathrm{g}}(z,r)=\frac{\Sigma(r)}{\sqrt{2\pi}H}
\exp\left(\;-z^2/2H^2\;\right)\;.
\end{equation}
In this expression the quantity $H$ denotes the pressure scale height
of the gas given by $H=c_{\mathrm{s}}/\Omega_{\mathrm{k}}$, where
$c_{\mathrm{s}}=\sqrt{kT/\mu}$ is the isothermal sound speed and
$\Omega_{\mathrm{k}}=\sqrt{GM_{\star}/r^3}$ denotes the Kepler
frequency. The quantities $k$ and $G$ are the Boltzman constant and
the gravitational constant, respectively. The mean molecular weight
$\mu$ is assumed to be 2.3 $m_{\mathrm{p}}$ (mixture of molecular
hydrogen and helium) where $m_{\mathrm{p}}$ is the mass of a
proton. 

With Eq.~(\ref{topf}), the gas mass density in the midplane of the
disk at 1~AU in our model with $\delta=0.8$ is given by
$10^{-11}$~g/cm$^3$. Adopting the MMSN model with $\delta=1.5$ leads
to a gas mass density of $4\times 10^{-10}$~g/cm$^3$ which is more
than one order of magnitude higher.

\subsection{Dust variables}

Before we introduce the vertical structure of the dust we define some
variables that describe the dust distribution. We define
$\rho_{\mathrm{d}}^{\mathrm{tot}}$ to be the total mass of the dust
per cm$^3$ at a certain point in space. To describe the particle mass
distribution we define a dust density $\rho_{\mathrm{d}}(m)$,
normalised such that
\begin{equation}
\rho_{\mathrm{d}}^{\mathrm{tot}}=\int_{0}^{\infty}
\rho_{\mathrm{d}}(m)\,\mathrm{d}m\;.
\end{equation}
Now, we introduce the number density $n$ by
\begin{equation}\label{snd}
n(m)=\frac{\rho_{\mathrm{d}}(m)}{m}\;.
\end{equation}
This quantity gives the number of solid particles of a certain mass
$m$ in a unity mass interval. Now, we can define the integrated number
density $w$ and the surface density of the dust by
\begin{equation}
\omega(m)=\int_{-\infty}^{+\infty} n(m)\,\mathrm{d}z
\end{equation}
and
\begin{equation}
\Sigma_{\mathrm{d}}(m)=m\omega(m)\;.
\end{equation}
To implement these expressions in a computer program we have to
introduce a mass grid \{$m_{\mathrm{k}}$\} and a measure \{$\Delta
m_{\mathrm{k}}$\}. With the definitions of the number density
$N_{\mathrm{k}}$ and the dust density $\rho_{\mathrm{k}}$ on the mass
grid
\begin{equation}\label{dataa}
N_{\mathrm{k}}=n(\bar{m}_{\mathrm{k}})\Delta m_{\mathrm{k}} \quad
\mbox{and} \quad
\rho_{\mathrm{k}}=\rho_{\mathrm{d}}(\bar{m}_{\mathrm{k}})\Delta
m_{\mathrm{k}}
\end{equation}
equation (\ref{snd}) implies
\begin{equation}
\rho_{\mathrm{k}}=m_{\mathrm{k}}N_{\mathrm{k}}\;.
\end{equation}
The quantity $n(\bar{m}_{\mathrm{k}})$ in Eq. (\ref{dataa}) is an
arbitrary value of the function $n$ within the interval $\Delta
m_{\mathrm{k}}$ around $m_{\mathrm{k}}$. To define a surface density
on a lattice we also have to introduce a vertical space grid
\{$z_{\mathrm{l}}$\} and its measure \{$\Delta z_{\mathrm{l}}$\}. The
surface density is then given by
\begin{equation}
\Sigma_{\mathrm{k}}=m_{\mathrm{k}}\sum_lN_{\mathrm{k}}(z_{\mathrm{l}})\,\Delta
z_{\mathrm{l}}.
\end{equation}

\subsection{Vertical structure of the dust}\label{verticalstructure}

In a protostellar disk the scale height of the dust is determined by
an equilibrium between two processes, namely the settling of the dust
towards the midplane of the disk due to vertical gravity, and the
vertical mixing of the dust due to turbulent diffusion. The more
turbulence there is, the harder it is for the dust to form a thin
layer since it is mixed up and transported back to the higher regions
of the disk.

To describe turbulence, we will use the $\alpha$-prescription of
\cite{ShaSun73}. In this prescription the turbulent diffusion coefficient of
the gas at a certain radius $r$ in the disk is parameterized by the scale
height of the gas $H$ and the isothermal sound speed $c_{\mathrm{s}}$ by
\begin{equation}
D_{\mathrm{g}}=\alpha c_{\mathrm{s}} H\;.
\end{equation}
The dimensionless parameter $\alpha$ determines the amount of turbulence in
the disk. Observations suggest a turbulent $\alpha$ value of $10^{-3}$
\citep{Hart98}. Numerical simulations of the magneto rotational instability
\citep{balhaw91} yield turbulence parameters of the same order of magnitude
\citep{Brand95,hgb95,sano04}. Moreover, \cite{Wei80} showed that there is a
minimal amount of turbulence in every protostellar disk corresponding to an
$\alpha$ value of about $10^{-6}$.

In addition to the $\alpha$-value we have to introduce a second dimensionless
number in order to describe the vertical structure of the dust. This so-called
Stokes number is defined by
\begin{equation}\label{stokesnumber}
\mathrm{St}_{\mathrm{k}}=\Omega_{\mathrm{k}}
\frac{a_{\mathrm{k}}\rho_{\mathrm{s}}}{c_{\mathrm{s}}\rho_{\mathrm{g}}}
\alpha^{2q-1}\;.
\end{equation}
The variable $a_{\mathrm{k}}$ and $\rho_{\mathrm{s}}$ denote the radius of the
dust particle of mass k and its material density, respectively. If the Stokes
number is much smaller than unity, then the dust particles are strongly
coupled to the gas.  In this case, the motions of the dust are basically the
motions of the gas and both components have the same behaviour with regard to
diffusion. If St exceeds unity, then the particles decouple from the gas and
are hardly influenced by the turbulent motions of the gas. The turbulence
parameter $q$ in Eq.~(\ref{stokesnumber}) determines whether turbulent
diffusion in the disk is realized by big turbulent eddies moving slow ($q=1$)
or by small turbulent eddies moving fast ($q=0$). Throughout this paper we
will assume that $q=1/2$ following \cite{cuzzi01} and \cite{schr04} unless
otherwise stated.

The dimensionless turbulence parameter $q$ is also connected with the
velocity $v_t$ of the large turbulent eddies,
\begin{equation}
v_t=\alpha^qc_{\mathrm{s}}\;,
\end{equation}
which significantly influences the relative turbulent velocities of
the dust (cf. Sec.~\ref{rv}) and, hence, its coagulation and
fragmentation time scales. Various authors have used quite different
values for $q$ during the past decades which led to very different
turbulent eddy velocities and, hence, different relative particle
velocities produced by turbulence. While \cite{morfill84},
\cite{weimess1} or \cite{weicuzzpp3} use turbulent gas velocities of
$\alpha c_{\mathrm{s}}$, which implies $q=1$, more recent publications
explicitly derived $q=1/2$ which leads to
$v_t=\sqrt{\alpha}c_{\mathrm{s}}$ \citep{dms95,cuzzi01,food}. If $q$
exceeds 1/2 then the time scale of the largest eddy becomes larger
than an orbital time scale since
$\tau_{\mathrm{eddy}}\sim\alpha^{1-2q}/\Omega_{\mathrm{k}}$. Turn over
frequencies smaller than the Kepler frequency appear unphysical to
us. Therefore, we follow \cite{cuzzi01} and adopt $q=1/2$.

With the two dimensionless numbers, $\alpha$ and St, the scale height of the
dust $h_{\mathrm{k}}$ of a certain grain mass $m_{\mathrm{k}}$ is given by
\citep{dms95,food,brauer07}
\begin{equation}\label{dsh}
\left(\frac{h_{\mathrm{k}}}{H}\right)^2=
\frac{\alpha}{\min(\mathrm{St_{\mathrm{k}}},1/2)(1+\mathrm{St_{\mathrm{k}}})}\;.
\end{equation}
Since the dust scale height $h_{\mathrm{k}}$ can not exceed the gas scale
height $H$ we restrict $h_{\mathrm{k}}$ to be at most $H$. With this last
expression the vertical structure of the dust particles with mass
$m_{\mathrm{k}}$ is given by
\begin{equation}\label{dustdens}
\rho_{\mathrm{k}}(r,z)=\frac{\Sigma_{\mathrm{k}}}{\sqrt{2\pi}h_{\mathrm{k}}}
\exp\left(\;-z^2/2h_{\mathrm{k}}^2\;\right)\;.
\end{equation}
In this equation $\Sigma_{\mathrm{k}}$ denotes the surface density of
the dust with mass $m_{\mathrm{k}}$.

\subsection{Radial motion of the dust}\label{drift}

In this section we will present the equations of radial motion for solid
particles. We first recapitulate the radial drift of individual particles and
particle motion due to gas accretion. After this we introduce the equations
for radial mixing due to turbulent diffusion. Finally, we will discuss the
continuity equation.

The following equations, which describe the radial drift of individual dust
particles, are valid as long as the dust-to-gas ratio $\epsilon$ does not
exceed unity. If, however, the dust density becomes larger than the gas
density then the dust starts to have a back-reaction on the motion of the gas
in a non negligible way. One possible scenario, where the dust-to-gas ratio
can exceed unity, is when the particles settles into a thin midplane layer due
to low turbulence in the disk. In this dense midplane layer these so-called
collective effects can become of importance and the radial drift equations
have to be modified in an appropriate way. This was discussed in detail by
\cite{NakSekHay86}.

\subsubsection{Radial drift of individual particles}\label{rdoip}

We consider two different sources for the radial drift of solid particles. The
first one is the radial drift of individual particles itself. The second
source is due to the accretion process of the gas.

The dust particles behave entirely independently and the gas is assumed not to
be affected by the dust at all. The crucial particle characteristics, that
determine the drift of solid particles, is the Stokes number introduced in the
last section. In other words, the coupling strength between the gas and the
dust determines the radial drift. With this quantity the radial drift of
individual dust particles of mass $m_{\mathrm{k}}$ is given by
\citep{Whipple72,Wei77}
\begin{equation}\label{pd1}
v_{\mathrm{dust,k}}=
-\frac{2v_{\mathrm{n}}}{\mathrm{St_{\mathrm{k}}}+\frac{1}{\mathrm{St_{\mathrm{k}}}}}\;.
\end{equation}
The radial drift of the dust is maximal for $\mathrm{St_{\mathrm{k}}}=1$. For
this reason, the radial drift barrier can be regarded as a region around
$\mathrm{St}_{\mathrm{k}}=1$. The maximum drift velocity $v_{\mathrm{n}}$ in
the last equation can be calculated according to
\begin{equation}\label{pd2}
v_{\mathrm{n}}=\frac{c_{\mathrm{s}}^2}{2V_{\mathrm{k}}}\left(
\delta+\frac{7}{4}\right)\;.
\end{equation}
With the numbers mentioned in the last section the last expression yields a
maximum drift velocity of 45 m/s at 1 AU. Since the power law index of the
temperature is $-1/2$ the maximum drift velocity $v_{\mathrm{N}}$ is
independent of the location in the disk. 

The second source for radial velocity of the dust is due to the
accretion process of the gas. In a gaseous disk angular momentum is
transported to the outer regions of the disk under the action of
viscous stress. This transport of angular momentum is connected with a
radial accretion velocity of the gas. This gas velocity also affects
the motions of the dust. Small particles are strongly coupled to
the motions of the gas and if the gas moves inwards then also the
small dust particles move inwards. For larger particles,
i.e. particles with Stokes number larger than unity, the motions of
the gas become less and less important since larger dust particles
decouple from the gas.

\cite{TakLin02} have calculated the radial accretion velocity of the gas due
to viscous stress,
\begin{equation}
v_{\mathrm{gas}}=-3\alpha\;\frac{c_{\mathrm{s}}^2}{V_{\mathrm{k}}}
\left(\frac{3}{2}-\delta\right).
\end{equation}
In the last expression we already adopted the temperature dependency $T\propto
r^{-1/2}$ according to eq.~(\ref{temp}). The quantity $V_{\mathrm{k}}$ denotes
the Kepler velocity $V_{\mathrm{k}}=\Omega_{\mathrm{k}}r$. With a turbulence
parameter of $\alpha=10^{-3}$ and the values that were already mentioned in
the last section the last equation yields a radial gas accretion velocity of 5
cm/s.

\begin{figure}
\begin{center}
\includegraphics[scale=0.5]{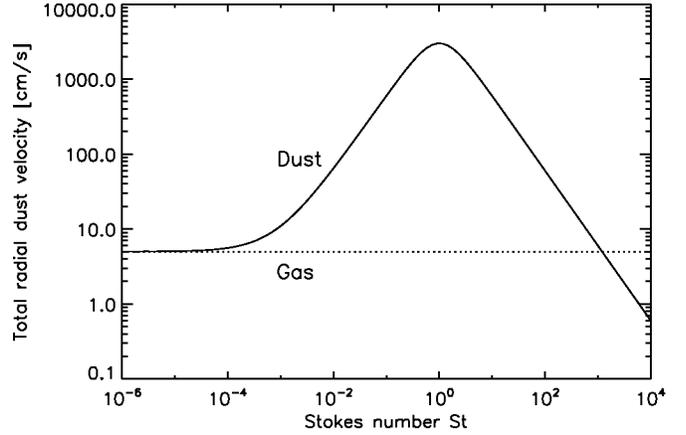}
\caption{The total inward radial velocity of a single dust particle as
  a function of Stokes number St (solid line) as discussed in
  Sec.~\ref{rdoip}. The dotted line denotes the accretion velocity of
  the gas. The turbulence parameter $\alpha$ is $10^{-3}$ in this
  calculation. \label{totaldrift}}
\end{center}
\end{figure}

Now, the total radial drift velocity of solid particles of mass
$m_{\mathrm{k}}$ is given by \citep{kornet01}
\begin{equation}\label{totdv}
v_{\mathrm{dust,k}}^{\mathrm{tot}}=v_{\mathrm{dust,k}}
+\frac{v_{\mathrm{gas}}}{1+\mathrm{St_{\mathrm{k}}}^2}\;.
\end{equation}
A plot of this total drift velocity is shown is Fig.~(\ref{totaldrift}).

\subsubsection{Radial mixing of the dust and the continuity equation}

Turbulence can be interpreted as a kind of diffusion.
The turbulent diffusion coefficient for the gas $D_{\mathrm{g}}$ was already
introduced in the last sections. The equivalent quantity for the dust of a
certain grain mass is given by \citep{voelk80,CuzDobCha93,schr04}
\begin{equation}
D_{\mathrm{d,k}}=\frac{D_{\mathrm{g}}}{1+\mathrm{St_{\mathrm{k}}}}\;.
\end{equation}
This relation was already implicitly used in the expression for the dust layer
thickness Eq.~(\ref{dsh}). For small particles, i.e. particles with
$\mathrm{St_{\mathrm{k}}}<1$, the diffusion coefficient for the dust matches
the diffusion coefficient for the gas. For Stokes numbers larger than unity
$D_{\mathrm{d,k}}$ decreases continuously since the particles more and more
decouple from the turbulent gas.

The continuity equation of the dust of a certain grain size including radial
drift and turbulent mixing reads
\begin{equation}
\dot{\Sigma}_{\mathrm{k}}+\frac{1}{r}\partial_r\left(rF_{\mathrm{k}}\right)=0\;.
\end{equation}
The dust mass flux $F_{\mathrm{k}}$ is given by
\begin{equation}
F_{\mathrm{k}}=\Sigma_{\mathrm{k}}v_{\mathrm{dust,k}}^{\mathrm{tot}}-
D_{\mathrm{d,k}}\Sigma_{\mathrm{g}}\partial_r
\left(\frac{\Sigma_{\mathrm{k}}}{\Sigma_{\mathrm{g}}}\right)\;.
\end{equation}
The first term on the right side is the mass flux for the radial drift of
individual particles discussed in the last section. The second term is the
mass flux due to turbulent diffusion (radial mixing). 
 
\subsection{Dust coagulation}\label{coagu}

Two particles of mass $m_{\mathrm{i}}$ and $m_{\mathrm{j}}$ in a
protostellar disk tend to have different particle velocities
$v_{\mathrm{i}}$ and $v_{\mathrm{j}}$ \citep{beck00}. For example,
micrometer-sized particles are carried away with the gas while larger
boulders, like meter-sized bodies, which are decoupled from the gas,
are hardly affected by any gas motion. Small particles have high
relative velocities due to Brownian motion even for equal-sized
pairs. Larger particles have not. These relative velocities $\Delta
v_{\mathrm{ij}}$ lead to occasional collisions. The number of
collisions per second between two particle species with number
densities $N_{\mathrm{i}}$ and $N_{\mathrm{j}}$ can be calculated to
be
\begin{equation}\label{crate}
\frac{\mathrm{collisions}}{\mathrm{s}}=\Delta v_{\mathrm{ij}}
\sigma_{\mathrm{ij}} N_{\mathrm{i}}N_{\mathrm{j}}\;.
\end{equation}
We make the simplification that we take for $\Delta v_{\mathrm{ij}}$ the
average relative velocity between the two dust particles. In principal, the
relative velocities have stochastic variations, but we ignore them here. We
assume that collisions lead to coagulation with a certain sticking probability
$p_{\mathrm{c}}$. In general this probability depends on various particle
parameters like the size of the particle, solid particle density, the
'fluffiness' of the particles \citep{BluWur99}. It also depends on the
relative particle velocity $\Delta v_{\mathrm{ij}}$. Small particles tend to
stick to each other up to high relative velocities \citep{domtie97} while
larger bodies show the tendency to fragment even for small relative velocities
\citep{Benz00}. This sticking probability will be discussed in more detail in
section (\ref{probs}).

With the collision rate Eq.~(\ref{crate}) we can calculate the number
of dust particles per second with mass $m_{\mathrm{i}}$ which
coagulate with any dust particle of any mass,
\begin{equation}
\dot{N}_{\mathrm{i}}^{\mathrm{Loss}}=\sum_{\mathrm{j}}
\Delta v_{\mathrm{ij}}
\sigma_{\mathrm{ij}} p_{\mathrm{c}} N_{\mathrm{i}}N_{\mathrm{j}}\;,
\end{equation}
which corresponds to the loss term for the number density of particles
with mass $m_{\mathrm{i}}$. The factor $\Delta
v_{\mathrm{ij}}\sigma_{\mathrm{ij}} p_{\mathrm{c}}$ is often called
the coagulation kernel. The gain term for particles with mass
$m_{\mathrm{i}}$ due to the coagulation of smaller particles with mass
$m_{\mathrm{k}}$ and $m_{\mathrm{j}}$ reads
\begin{equation}
\dot{N}_{\mathrm{i}}^{\mathrm{Gain}}=
\sum_{m_{\mathrm{i}}=m_{\mathrm{k}}+m_{\mathrm{j}}}
\Delta v_{\mathrm{kj}}
\sigma_{\mathrm{kj}} p_{\mathrm{c}} N_{\mathrm{k}}N_{\mathrm{j}}\;. 
\end{equation}
Now we can introduce the full coagulation equation which is given by
\citep{smo16}
\begin{equation}\label{ce}
\dot{N}_{\mathrm{i}}=\dot{N}_{\mathrm{i}}^{\mathrm{Gain}}
-\dot{N}_{\mathrm{i}}^{\mathrm{Loss}}\;.
\end{equation}
The continuous formulation of the coagulation equation, as discussed by
\cite{saf69}, corresponds to a non-linear integro-differential equation.
However, this equation is rather difficult to solve both in its discrete or
continuous version. This is only within the realms of possibility for very
simple (and unfortunately unphysical) kernels. Therefore, we will solve the
coagulation equation numerically. The algorithm we will make use of is
described in detail in Appendix \ref{alg}.

\subsubsection{Relative dust particle velocities}\label{rv}

We will consider four different sources for relative particle
velocities which lead to coagulation: Brownian motion, differential
settling, turbulence and radial drift.

First, let us focus on Brownian motion. Two particles of mass $m_1$ and $m_2$
in a region of the disk with temperature $T$ have an average statistical
relative velocity due to Brownian motion given by
\begin{equation}
\Delta v_{\mathrm{B}}=\sqrt{\frac{8kT(m_1+m_2)}{\pi m_1m_2}}\;.
\end{equation}
This expression shows that relative thermal velocities are higher for smaller
dust particles than for larger dust particles. Hence, the growth process due
to Brownian motion is more effective for small particles than for large
particles. For example, if we assume a temperature of 200 K, a solid particle
density of 1 g/cm$^3$ and micrometer-sized particles then the relative
particle velocity due to Brownian motion is 0.2 cm/s. Particles of centimeter
in size lead to a relative velocity of $10^{-7}$ cm/s. This particular example
shows that there is practically no coagulation due to Brownian motion for
particles much larger than micrometer size. In general, growth by Brownian
motion leads to fractal structures and 'fluffy' aggregates (cluster-cluster
aggregates) \citep{oss93,kempf99}. However, we will ignore these intrinsic
properties of the dust particles in the course of this paper and assume a
constant solid material density. See, however, \cite{schmitt97} or
\cite{ormel07_2} for dust particle coagulation models including porosity at a
fixed radius in the disk.

Differential settling is the second process that leads to relative
velocities.  If we assume that the solid particles are smaller than
the mean free path of the gas then the equilibrium settling velocity
is given by $z\mathrm{St}\Omega_{\mathrm{k}}$ \citep{DD04}.  In this
expression St is the Stokes number introduced in section
\ref{verticalstructure}. However, for Stokes numbers larger than
unity, the equilibrium settling velocity model loses validity. Very
large bodies ($\mathrm{St}\rightarrow\infty$) above or below the
midplane follow an orbit that is tilted with respect to the
midplane. The settling velocity towards the midplane can not exceed
the vertically projected Kepler velocity $zV_{\mathrm{k}}/r$
corresponding to this inclined orbit. For this reason we restrict the
settling velocity to be the projected Kepler velocity at
most. Considering this, we adopt the following settling velocity in
our model,
\begin{equation}
v_{\mathrm{S}}=
\frac{z\mathrm{St}\Omega_{\mathrm{k}}}{1+\mathrm{St}}\;.
\end{equation} 
The relative settling velocity between two particles of mass $m_{\mathrm{i}}$
and $m_{\mathrm{j}}$ at a height $z$ above the midplane then reads 
\begin{equation}
\Delta
v_{\mathrm{S}}=z\Omega_{\mathrm{k}}
\left|\frac{\mathrm{St}_{\mathrm{i}}}{1+\mathrm{St}_{\mathrm{i}}}
-\frac{\mathrm{St}_{\mathrm{j}}}{1+\mathrm{St}_{\mathrm{j}}}\right|\;.
\end{equation}

The third source for relative velocities of particles in the disk is the
radial drift which was discussed in detail in section \ref{drift}. The
relative velocity in this case is simply the difference in the drift
velocities
\begin{equation}
\Delta v_{\mathrm{D}}=
|v_{\mathrm{dust,i}}^{\mathrm{tot}}-v_{\mathrm{dust,j}}^{\mathrm{tot}}|\;.
\end{equation}

The fourth relative velocity between the particles is due to
turbulence in the disk. Relative particle velocities produced by
turbulence were calculated numerically by \cite{voelk80} and
\cite{mizuno88}. \cite{wei84} fitted these results with analytical
formulas. Current work by \cite{ormel07} shows that these expressions
underestimate the turbulent relative velocities for particles with
large Stokes numbers. In this paper, we will use the expressions
calculated by \cite{ormel07}.

\begin{figure}
\begin{center}
\includegraphics[scale=0.5]{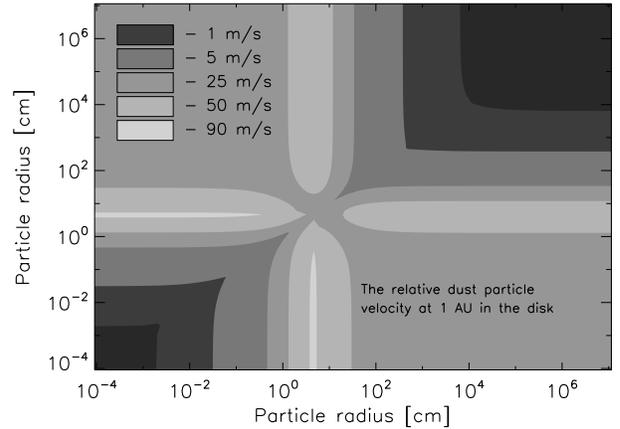}
\caption{Relative velocities of dust particles at 1~AU in the disk as
  discussed in Sec.~\ref{rv}. This calculation includes Brownian
  motion, differential settling and relative turbulent velocities. In
  this calculation we adopted a turbulent $\alpha$ value of
  $10^{-3}$. \label{vrel}}
\end{center}
\end{figure}

\begin{figure}
\begin{center}
\includegraphics[scale=0.5]{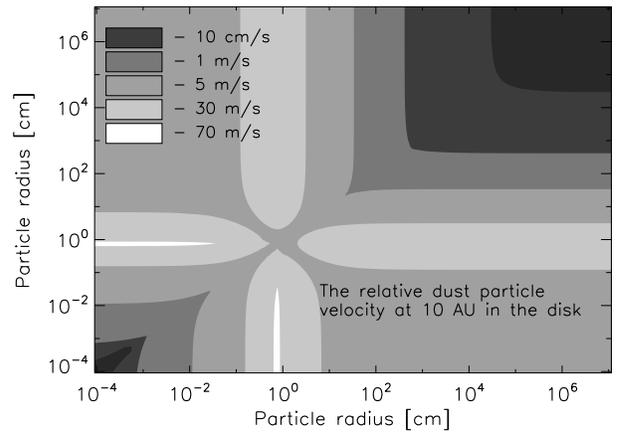}
\caption{As Fig.~\ref{vrel}, but now at 10 AU in the disk.\label{vrel2}}
\end{center}
\end{figure}

To give an impression of relative dust particle velocities in a protostellar
disk, Fig.~\ref{vrel} shows a contour plot of this quantity at 1~AU including
Brownian motion, differential settling, relative turbulent velocities and
relative particle drift velocities.  The same calculation at 10 AU is shown in
Fig.~\ref{vrel2}.


\subsection{Particle fragmentation}\label{fragmodel}

Collisions between particle aggregates do not necessarily lead to particle
growth. For sufficiently high relative collision velocities, the aggregates
may fragment into smaller bodies. The critical threshold velocity for this
destructive process generally depends on the mass of the colliding
particles. More precisely, fragmentation tends to play a non-negligible role
if the kinetic collisional energy of the particles is of the order of the
internal binding energy of the particles \citep{bordwe95}. Fragmentation
velocities of aggregates are usually of the order of a few cm/s up to several
10 m/s. While smaller particles tend to stick to each other up to high
relative particle velocities \citep{domtie97} larger bodies show the tendency
to fragment even for small relative velocities \citep{Benz00}. For simplicity,
we will assume a fixed threshold velocity for particle destruction
$v_{\mathrm{f}}$ which does not depend on the mass of the particles. However,
we will investigate how the results of the simulations change if
$v_{\mathrm{f}}$ is varied over a wide parameter range. The dependency of
$v_{\mathrm{f}}$ on the particle mass will be investigated in the near future
including laboratory results of dust particle collisions.

The result of destructive collisions between solid particles, i.e. the exact
particle distribution after fragmentation, is still a matter of
debate. Usually this particle distribution is described by a power-law,
\begin{equation}\label{fragres}
n(m)\,\mathrm{d}m\,\propto m^{-\xi}\mathrm{d}m\;.
\end{equation}
In this expression $n(m)\mathrm{d}m$ is the number of particles per unit
volume within the mass range $[m,m+\mathrm{d}m]$. The last decades involved
various attempts to determine the fragmentation parameter $\xi$. \cite{mat77}
and also \cite{dralee84} showed that the extinction and scattering of
starlight by interstellar dust can be reproduced by a power-law dependency
following $\xi=1.83$. Experimental studies found values for $\xi$ ranging
between 1.3 (low-velocity impacts) and 2 (catastrophic impacts)
\citep{davrya90,bm93}. Steady state solutions between coagulation and
fragmentation lead to $\xi=1.83$ as shown by \cite{doh69}. More recently,
\cite{taninanak96} argued that the very general result $\xi=1.83$ is a direct
implication of the self-similarity of the particle size distribution. In this
paper we will assume the $\xi$-value 1.83 if not otherwise noted.
    
The process of fragmentation between particles which have the same
mass is different from the fragmentation of particles whose mass
differ by orders of magnitude. Two bodies of equal mass may destroy
each other. Small dust grains, however, are not able to destruct a
meter-sized body. But they can excavate a small crater in the larger
target. This process is usually called 'cratering'. We will assume
that cratering sets in if the mass of the colliding bodies differs by
more than one order of magnitude. In this case, the smaller dust
particle $m_s$ excavates a crater which contains a factor $\chi$ times
its own mass, i.e. $m_{\mathrm{crater}}=\chi m_s$. The parameter
$\chi$ is set to unity if not otherwise noted.  The mass of the
smaller body and the crater ejecta are then redistributed according to
Eq.~(\ref{fragres}). On the other hand, if the mass of the colliding
particles differs by less than an order of magnitude, i.e. in the
non-cratering case, then the total mass is redistributed following
Eq.~(\ref{fragres}).

\begin{figure}
\begin{center}
\includegraphics[scale=0.5]{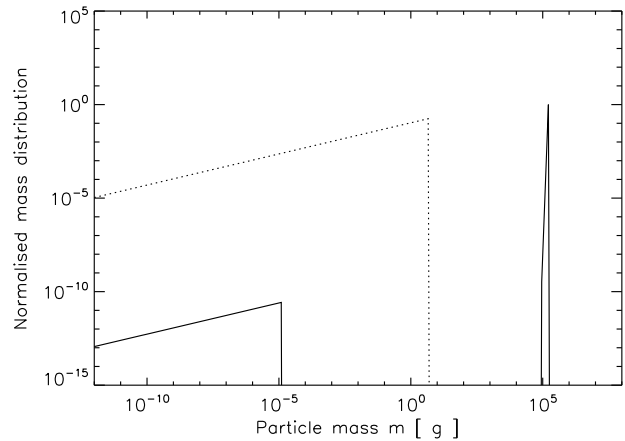}
\caption{The assumed fragmentation results of collisional destruction as
  discussed in Sec.~\ref{fragmodel}. The solid line shows an example for
  cratering. The larger body has a mass of $\sim 10^5$~g. The smaller dust
  grain, which is destroyed in this process, had a mass of $\sim
  10^{-5}$~g. The dotted line corresponds to the collision of two particles
  with the same mass. In this calculation the fragmentation parameter $\xi$ is
  assumed to be 1.83 which means that most of the fragmentation results are at
  the large end of the particle size distribution.
 \label{frag}}
\end{center}
\end{figure}

To illustrate the results of fragmentation Fig.~\ref{frag} shows the outcome
of a destructive collision as modeled in the paper at hand. The solid line
shows the outcome of fragmentation in the case of cratering. The dotted line
corresponds to the fragmentation results of two particles with the same mass.

\subsection{Coagulation and fragmentation probabilities}\label{probs}

If the collision velocity of two boulders is sufficiently large then the
particles tend to fragment into smaller bodies instead of coagulating to
larger aggregates. We will assume that the probability for fragmentation
$p_{\mathrm{f}}$ only depends on the relative particle velocity $\Delta v$ and
adopt the following expression for this probability,
\begin{equation}\label{fproba}
p_{\mathrm{f}}(\Delta v)=\left(\frac{\Delta v}{v_{\mathrm{f}}}\right)^{\psi}\Theta\left(v_{\mathrm{f}}-v\right)+\Theta\left(v-v_{\mathrm{f}}\right)\;.
\end{equation}
The two Heaviside step functions $\Theta$ ensure that the particles fragment
with 100\% probability if the relative particle velocity $\Delta v$ is larger
than the critical fragmentation velocity $v_{\mathrm{f}}$. For $\Delta
v<v_{\mathrm{f}}$ we assume that there is always a possibility for
fragmentation given by $(v/v_{\mathrm{f}})^{\psi}$. We will investigate the
influence of the critical fragmentation velocity $v_{\mathrm{f}}$ and the
index $\psi$. The value of $\psi$ is set to unity if not otherwise noted. The
probability for coagulation $p_{\mathrm{c}}$ is given by
$p_{\mathrm{c}}=1-p_{\mathrm{f}}$. The last expression implies that the
particles either coagulate or they fragment. We do not allow the particles to
collide and not to undergo either the process of coagulation or
fragmentation. However, just for the moment let us assume that
$p_{\mathrm{f}}+p_{\mathrm{c}}<1$. If this last expression holds then the time
scales for coagulation and fragmentation increase. However, this issue might
be considered in a forthcoming paper.

\section{Simulation results}

In the following we will present the results of our numerical
simulations. These simulations include various effects, for example,
different particle growth mechanisms (Brownian motion, differential
settling, etc.), the radial drift of the dust and particle
fragmentation. To illustrate the influence of these effects on the
particle growth, we will proceed in certain steps. In every step more
and more effects will be included. In the first step we will consider
the growth of the dust particles at various radii in the disk. In this
part we do not allow the particles to move radially. In the second
step, however, we will also include the radial motion of the dust
which was discussed in detail in section \ref{drift}. In the last step
we will additionally consider fragmentation.

\subsection{Step 1 - Coagulation only}\label{s1}

What are the growth time scales of the solid particles at different
radii in the disk? To answer this question, we will not allow any
radial motion of the particles. We glue the dust to a certain radial
position even though the radial drift of the dust is potentially very
high. We also do not allow particle fragmentation. The coagulation of
solid particles at a fixed radius in the disk was for example also
treated by \cite{schmitt97} and \cite{DD05}, \cite{naka81},
\cite{tan05} and recently \cite{ciesla2007}. We assume that the mass
of the disk is 1\% of the central mass, an initial dust-to-gas ratio
of $\epsilon_0=10^{-2}$ and a solid material density\footnote{10\%
  silicate, 30\% carbonaceous material and 60\% ice} of
$\rho_{\mathrm{s}}=1.6$~g/cm$^3$. The turbulent $\alpha$ parameter is
$10^{-3}$ and the turbulent $q$ parameter is set to be $1/2$. At the
beginning of each simulation the dust is equally distributed between a
dust particle size of 0.5 $\mu$m and 0.8 $\mu$m.

\begin{figure}
\begin{center}
\includegraphics[scale=0.74]{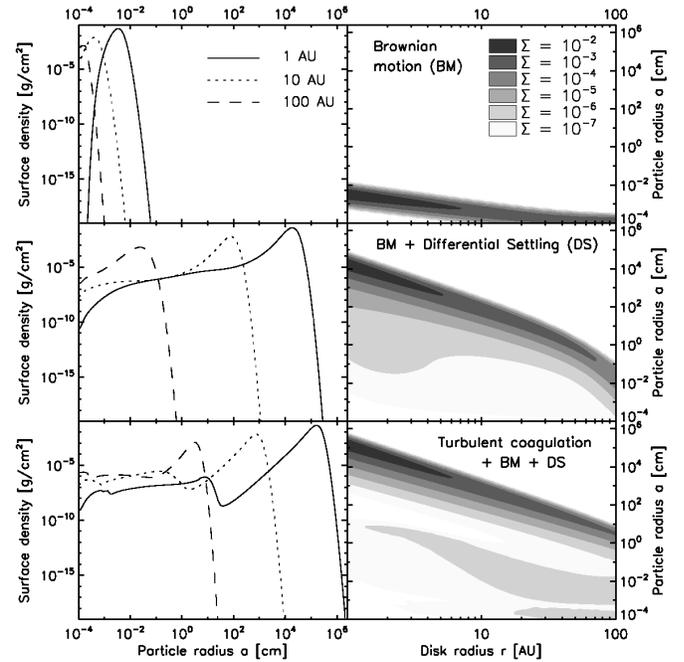}
\caption{These plots show the particle size distribution at different
  radii in the disk after 1 Myrs of disk evolution as discussed in
  Sec.~\ref{s1}. The left and the right plot always belong
  together. From top to bottom more and more growth mechanisms are
  included in the simulations. The upper panel shows coagulation only
  due to Brownian motion (BM). The second panel shows BM and
  differential settling (DS). Finally, BM + DS and turbulent
  coagulation (TC) are shown in the lowest panel. The left plots show
  the surface density of the dust at 1, 10 and 100 AU as a function of
  particle radius. On the right side the corresponding contour plots
  of the dust surface density are shown as a function of the radial
  location in the disk and the particle radius. In these simulations
  the radial drift as well as the fragmentation of the dust particles
  were neglected.
  \label{coag}}
\end{center}
\end{figure} 
 
Let us first focus on the particle growth due to Brownian motion at
different radii in the disk. The result of this simulation, i.e. the
particle size distribution after 1 Myrs, is shown in the upper panel
of Fig.~\ref{coag}. According to these results dust particles grow
from sub-micrometer to $\sim 30$~$\mu$m in radius in 1 Myrs at 1 AU in
the disk. At 10 AU the particle distribution has a maximum for $a\sim
4$~$\mu$m. At 100 AU most of the dust is roughly a micrometer in
size. We conclude that particle growth due to Brownian motion is not
very effective, which is a well known result
\citep{oss93,schmitt97,DD05}. However, Brownian motion is an important
effect for the following reason. We calculate the relative velocities
due to Browian motion, differential settling and turbulence for
$a=0.6$~$\mu$m equal-sized particles at 1 AU in the disk. While the
relative particle velocity due to Brownian motion is 0.4 cm/s the
relative turbulent velocity is in the order of $10^{-8}$ cm/s. The
relative velocity due to differential settling is practically
zero. Dust particle growth due to differential settling or turbulence
gets of importance only for larger particles. Therefore, Brownian
motion is a trigger mechanism for the entire coagulation process which
was noted before by \cite{wei84}.

Now, we will additionally include coagulation due to differential settling
into our model. The result of this simulation is shown in the second panel of
Fig.~\ref{coag}. This plot shows that particles have grown to more than $10^4$
cm in radius at 1 AU in the disk after 1 Myrs. This particle size is more than
6 orders of magnitude larger than the grain size after 1 Myrs caused by
Brownian motion. Most particles at 10 AU and 100 AU have grown to sizes of
about 1~m and 100~$\mu$m, respectively. We conclude that differential settling
is an effective growth mechanism which can create large boulders in the inner
parts of the disk. Note that in our model the vertical mixing continuously
allows the grains to go back up again and grow again by differential
settling. Therefore, the maximal size formula of \cite{saf69} does not apply
here.

Apart from the fact that particles grow to much larger sizes if
differential settling is included, Fig.~\ref{coag} also shows that
there is always a certain amount of small particles that remains in
the disk and that does not coagulate for at least 1 Myrs. After this
time, roughly 6\% of the dust between 1 and 75~AU is still present in
grains $<1$~mm. The reason for this is the following. Not all of the
dust particles coagulate at the same time. While a certain fraction of
the dust has already grown to larger sizes and formed a thinner dust
layer according to Eq.~(\ref{dsh}), a certain fraction of small dust
remains in the higher regions above the midplane. These small dust
particles high above the midplane are subject to a rather slow
coagulation process. The dust densities above the midplane are low
after most of the dust already settled closer to the midplane. This
leads to long growth time scales according to
Eq.~(\ref{crate}). Larger particles close to the midplane can not
sweep up the smaller particles above the midplane since turbulence is
not able to stir them up so far. For this reason small particles
remain in the disk for a long time.

We will now also include relative velocities of the particles caused
by random turbulent motions. The result of this simulation is shown in
Fig.~\ref{coag} in the lower panel. This plot indicates that the
dominant grain size at 1 AU, i.e. the grain size corresponding to the
surface density maximum, changes by a factor of $\sim 10$ if turbulent
coagulation is included in the simulation. The dominant particle
radius at 1~AU is $\sim 10^5$~cm. At 100~AU, random turbulent motions
also speed up the coagulation process which leads to particles of a
few centimeters in radius after 1 Myrs of disk evolution. Without
relative turbulent velocities included in the simulation, the particle
radius was two orders of magnitude smaller.

\subsection{Step 2 - Coagulation and radial motion}\label{step2}

We will now include radial motion, both as transport and as extra source of
relative velocities for coagulation. This significantly changes the results of
the last section. We find that the radial drift of solid particles is so high
that the dust drift into the evaporation zone long before larger particles in
the disk can possibly form. This happens even though an additional source for
coagulation is introduced which decreases the coagulation time scales. We will
investigate if particles can in some way ''break through'' the radial drift
barrier.

\subsubsection{Time evolution of the disk}\label{sze}

\begin{figure}
\begin{center}
\includegraphics[scale=0.73]{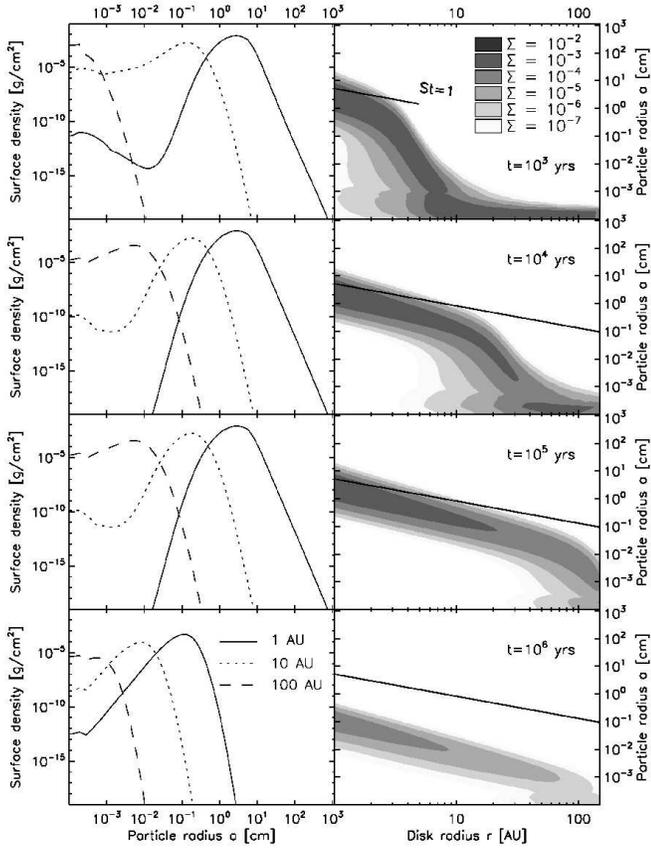}
\caption{The particle size distribution at different radii in the disk at
  different times of disk evolution as discussed in Sec.~\ref{sze}. In this
  simulation all particle gowth mechanisms are included as well as the radial
  motion of the dust. The fragmentation of particles is neglected. The left
  and the right plots always belong together. The left column shows the
  surface density as a function of particle radius at 1, 10 and 100 AU. The
  right column shows the corresponding contour plots of the surface density as
  a function of disk radius and particle radius. The white lines in the
  contour plots denotes the particle radius for which the Stokes number is
  unity (i.e. largest radial drift and largest radial velocities).
  \label{cdrift}}
\end{center}
\end{figure}

Fig.~\ref{cdrift} shows the time evolution of the model. This plot indicates
that cm-dm-sized particles form in the inner regions of the disk ($<2$ AU)
within the first $10^3$ yrs. Compared to the outer parts of the disk the
formation of these particles appears rather quickly due to comparatively high
gas and dust densities and high temperatures. With increasing distance from
the central star the formation of larger particles gets more and more
difficult. At 10~AU in the disk, it is still possible to form mm-sized
particles in $10^4$ yrs according to Fig.~\ref{cdrift}. However, in the outer
parts of the disk ($>100$~AU) the dominant particle size of the dust never
exceeds 0.1~mm at any time. The disregard of radial drift in the previous
section led to particle sizes of more than a centimeter at 100 AU after
1~Myrs, which is orders of magnitude larger.

The neglect of radial drift, as discussed in section \ref{s1},
involved a permanent amount of small particles which was present
throughout the disk for at least 1 Myrs. These small particles were
located high above the midplane and were subject to a rather slow
coagulation process due to relatively low dust
densities. Fig.~\ref{cdrift} indicates that there is a smaller
remaining amount of small dust if radial motion is taken into
account. This is due to the following reason. Even the small particles
in the higher regions of the disk can have relative radial velocities
of the order of some mm/s or even cm/s.  These higher relative
velocites lead to higher collision rates and, hence, to a depletion of
the small dust grains.

After $10^5$ yrs of disk evolution, the average particle size at a certain
radius in the disk starts to decrease in time. To give an example, after
$10^5$ yrs the dominant dust grain radius at 1~AU is $\sim 1$~cm. After 1~Myrs
this value is about an order of magnitude lower. While particles drift inward
from a certain radial position they are replaced by other particles from the
outer parts of the disk. The coagulation time scales are larger in the outer
parts of the disk which means that particles grow to smaller sizes in the same
time. Therefore, the particles that reach a certain position are smaller than
the particles that drift away and, hence, the average dust particle size
decreases.

In the simulation shown in Fig.~\ref{cdrift} the Stokes number of the dominant
particles never exceeds unity, i.e. never breaks the radial drift barrier, at
any disk radius considered at any time. This is indicated by the $\mbox{St}=1$
line which is also shown in this plot. At $\sim 1$~AU in the disk, the
simulation shows that particles may grow to sizes that correspond to a Stokes
number slighly smaller than unity. In the following we will investigate if the
particles may break through the $\mbox{St}=1$ barrier for certain disk
parameters.

\subsubsection{Effect of disk mass}\label{secmass}

\begin{figure}
\begin{center}
\includegraphics[scale=.5]{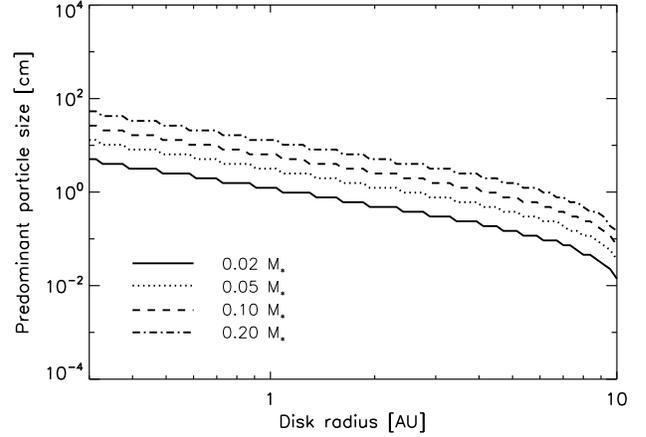}
\caption{The effect of disk mass on the particle growth as discussed in
  Sec.~\ref{secmass}. Shown is the dominant dust particles radius after $10^4$
  yrs of disk evolution for different disk masses between 0.2 and 10 AU. The
  turbulent $\alpha$ parameter is $10^{-3}$ and the initial dust-to-gas ratio
  is $10^{-2}$.
  \label{diskmass}}
\end{center}
\end{figure}

We investigate the effect of disk mass on the particle growth. The result of
this investigation can be seen in Fig.~\ref{diskmass}. This plot shows the
dominant dust particle size for different disk masses after $10^4$ yrs of disk
evolution as a function of disk radius. We find that the particle size
increases by an order of magnitude if the disk mass is increased from 1\% to
20\% of the central mass. Larger disk masses lead to higher gas and dust
densities and, hence, to higher collision rates according to
Eq.~(\ref{crate}). Therefore, dust particles can grow to larger sizes over the
same time interval.

The Stokes number of the dominant particles is always smaller than
unity. Of course, particles may grow to larger sizes which increases
the Stokes number since $\mbox{St}\propto a$. However, larger disk
masses also lead to higher gas densities which again decreases the
Stokes number because $\mbox{St}\propto 1/\rho_{\mathrm{g}}$.
Finally, both effects cancel out and the disk mass seems to plays a
minor role in breaking the radial drift barrier.

\subsubsection{Effect of turbulence}\label{secturb}

\begin{figure}
\begin{center}
\includegraphics[scale=.5]{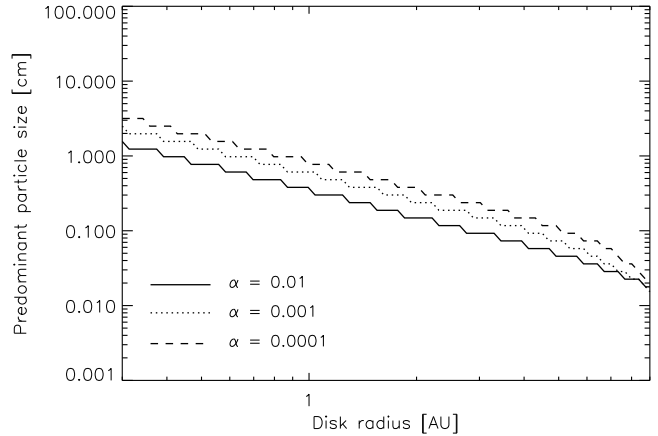}
\caption{Same plot as Fig.~\ref{diskmass} but now showing the effect of
  turbulence on the particle growth as discussed in Sec.~\ref{secturb}. Shown
  is the dominant particles size after $10^4$ yrs of disk evolution for
  different turbulent $\alpha$ parameters between 0.2 and 10 AU. The disk mass
  is $10^{-2}$ $M_{\star}$ and the initial dust-to-gas ratio is $10^{-2}$.
  \label{alpha}}
\end{center}
\end{figure} 

As in the last section, we calculate the dominant particle size after
$10^4$ years but now for different turbulent $\alpha$-parameters
instead of different disk masses. The initial dust-to-gas ratio in
this simulation is $10^{-2}$, the disk mass is $10^{-2}$ $M_{\star}$
and the result is shown in Fig.~\ref{alpha}.

One would intuitively think that in a certain time particles can grow
to larger sizes in highly turbulent disks than in low-turbulent
disks. Fig.~\ref{alpha} shows, however, that the dominant particle
size after $10^{4}$ yrs is only weakly dependent on the turbulence
parameter $\alpha$. If $\alpha$ changes by two orders of magnitude
then the dominant particle size only changes by a factor of two. This
can be understood by the following consideration.

A high amount of turbulence in the disk leads to high relative turbulent
particle velocities \citep{voelk80,wei84,cuzzi01}. These high relative
velocities cause high collision rates, cf. Eq.~(\ref{crate}), which favour the
process of coagulation. For this reason particles should have grown to larger
sizes in highly turbulent disks. On the other hand, a large amount of
turbulence leads to thick particle layers. The dust is stirred up in the
higher regions of the disk which causes the average dust densities to
decrease. The collision rates in Eq.~(\ref{crate}) are proportional to the
particle number densities.  Lower dust particle densities lead to longer
coagulation time scales.

The two determining factors for the growth time scales, the relative turbulent
particle velocity and the dust density, seem to cancel out if the amount of
turbulence in the disk is varied. Hence, different $\alpha$-parameters lead to
the same particle size over the same time interval.

\subsubsection{Effect of the initial dust-to-gas ratio}\label{seceps}

\begin{figure}
\begin{center}
\includegraphics[scale=.74]{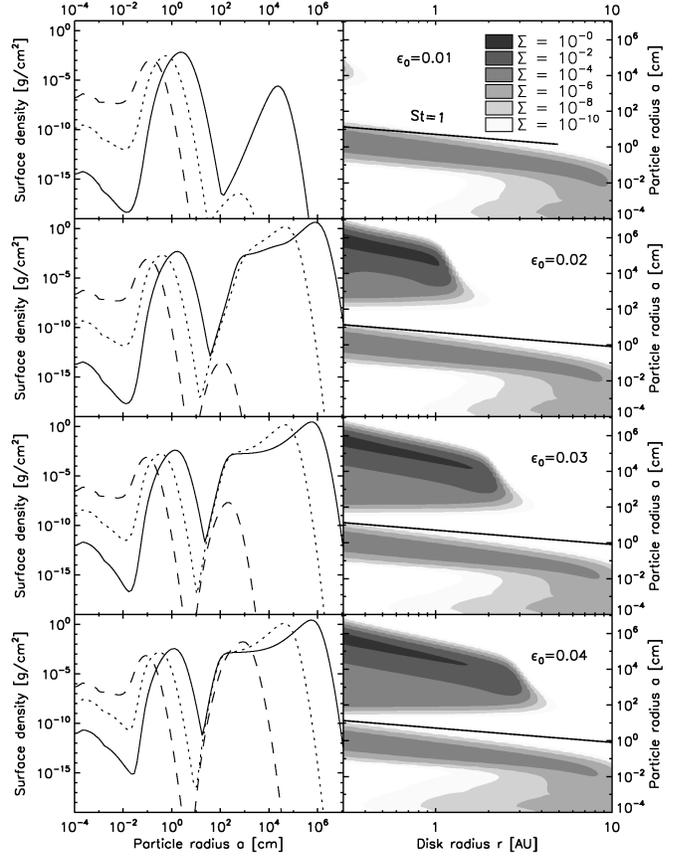}
\caption{These plots shows the effect of the initial dust-to-gas ratio
  on the particle growth as discussed in Sec.~\ref{seceps}. The right
  side shows contour plots of the surface density as a function of
  disk location and particle radius for 4 different initial
  dust-to-gas ratios after $10^4$~yrs of disk evolution. The
  corresponding left plots show the surface density as a function of
  particle radius for 3 different locations in the disk (0.3~AU -
  solid, 1~AU - dotted, 3~AU - dashed) after the same time. The disk
  mass is $10^{-2}$ $M_{\star}$ and the turbulent $\alpha$ parameter
  is $10^{-3}$. For initial dust-to-gas ratios which are slighly
  higher than 1\% the particles break through the radial drift
  barrier.
  \label{eps}}
\end{center}
\end{figure}

We now investigate the effect of the initial dust-to-gas ratio on the growth
time scales and the particle size distribution. We consider a disk mass of
$10^{-2}$ $M_{\star}$ and a turbulence parameter $\alpha$ of $10^{-3}$. The
result of this investigation is shown in Fig.~\ref{eps}.

\begin{figure}
\begin{center}
\includegraphics[scale=0.74]{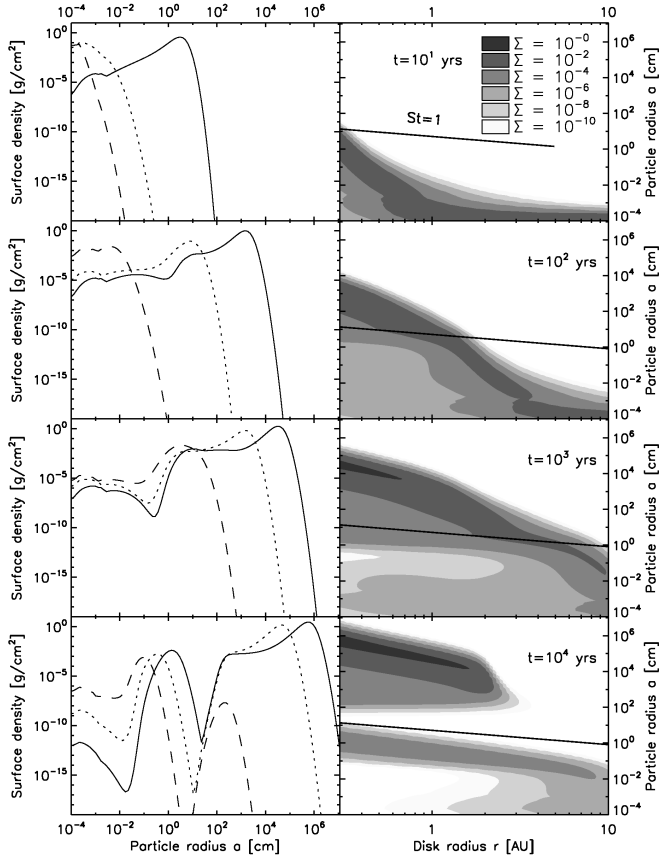}
\caption{This plot shows the results of a simulation in which the particles
  can break through the radial drift barrier as discussed in
  Sec.~\ref{seceps}. Particle fragmentation is neglected in this
  simulation. Shown is the surface density distribution for the first
  $10^4$~yrs of disk evolution for an initial dust-to-gas ratio of 0.03 as a
  function of disk radius and particle radius. The disk mass is $10^{-2}$
  $M_{\star}$ and the turbulent $\alpha$ parameter is $10^{-3}$. The right
  side is a contour plot of the surface density. The left side shows the
  absolute values of the surface density for 3 different disk radii (solid -
  0.3~AU, dotted - 1~AU, dashed - 3~AU).
  \label{eps2}}
\end{center}
\end{figure}

This contour plot shows the surface density of the particle
distribution as a function of disk location and particle radius for
four different initial dust-to-gas ratios after $10^4$~yrs of disk
evolution. These results indicate that $10^4-10^5$~cm sized boulders
can form in the inner parts of the disk ($<3$ AU) subject to the
condition that the initial dust-to-gas ratio of the disk is higher
than 1\%. This means that the dust particles may overcome the radial
drift barrier if the dust-to-gas ratio is slighly higher than usually
assumed. A contour plot of the surface density distribution with
$\epsilon_0=0.03$, i.e. in the case where the particles are able to
break through the radial drift barrier for disk radii $<3$ AU, as a
function of time is shown in Fig.~\ref{eps2}.

To understand this importance of the initial dust-to-gas ratio we
consider the growth rate of the dust particles as given by
\cite{kornet01},
\begin{equation}\label{kornetf}
\dot{a}=\frac{\rho_{\mathrm{d}}}{\rho_{\mathrm{s}}}\Delta v\;.
\end{equation}
If we assume that the particles have Stokes numbers smaller than unity then
the relative turbulent particle velocity is given by
\citep{cuzzi01,weicuzzpp3}
\begin{equation}\label{relturbvel}
\Delta
v_{\mathrm{turb}}\propto\sqrt{\alpha\mathrm{St}}\,c_{\mathrm{s}}\;.
\end{equation}
The dust mass density can be approximated by
  $\rho_{\mathrm{d}}\propto\Sigma_{\mathrm{d}}/h$ so that we obtain
\begin{equation}
\dot{a}=\frac{1}{\rho_{\mathrm{s}}}
\frac{\Sigma_{\mathrm{d}}}{h}\sqrt{\alpha\mathrm{St}}\,c_{\mathrm{s}}\;.
\end{equation}
With the height of the dust layer Eq.~(\ref{dsh}), the last expression
can be written as (for $\mathrm{St}<1$)
\begin{equation}\label{cts}
\dot{a}=\frac{1}{\rho_{\mathrm{s}}}\epsilon_0\Sigma_{\mathrm{g}}\mathrm{St}\Omega_{\mathrm{k}}\;.
\end{equation}
If we also take into account the definition of the Stokes number in
Eq.~(\ref{stokesnumber}), then most quantities cancel each other out,
particularly the gas surface density $\Sigma_{\mathrm{g}}$, leading to
\begin{equation}
\dot{a}=a\epsilon_0\Omega_{\mathrm{k}}
\end{equation}
with the solution
\begin{equation}\label{solu}
a=a_0e^{\epsilon_0\Omega_{\mathrm{k}}t}\;.
\end{equation}
This expression shows that only the initial dust-to-gas ratio
$\epsilon_0$ and the Kepler frequency $\Omega_{\mathrm{k}}$ determine
the turbulent growth time scales as long as $\mathrm{St}<1$. According
to Eq.~(\ref{solu}), the time scales are not linear dependent on the
initial dust-to-gas ratio. An increase of $\epsilon_0$ leads to an
exponential decrease of the growth time scales. This strong dependency
unveils the crucial importance of this initial
parameter. Eq.~(\ref{solu}) also shows that turbulent coagulation
occurs faster in the inner parts of the disk than in the outer parts
since $\Omega_{\mathrm{k}}\propto r^{-1.5}$. For this reason, the
particles first break through the radial drift barrier in the inner
parts of the disk (cf. Fig.~\ref{eps}).

In Section \ref{secmass} and \ref{secturb} we have seen that the
dominant particle size only shows a weak dependency on the disk mass
and the amount of turbulence in the disk. This can also be explained
by Eq.~(\ref{cts}). The turbulent growth rate of the dust is neither
dependent on the disk mass nor on the turbulent $\alpha$
parameter. Moreover, this expression also indicates that the disk
temperature and intrinsic particle properties like solid density are
rather unimportant as long as the Stokes number of the particles is
smaller than unity and turbulence is the leading process that triggers
coagulation. 

However, \cite{ormel07_2} have shown that the porosity of dust
particles actually matters in the early phases of disk evolution. This
discrepancy is due to the fact that the Eq.~(\ref{solu}) only holds if
$\mathrm{St}>\alpha$ while \cite{ormel07_2} considered particles with
$\mathrm{St}<\alpha$. Moreover, Brownian motion is the main source for
relative particle velocities for small dust grains in the early disk
evolution while the derivation of Eq.~(\ref{solu}) assumes that
turbulence is the major source for relative particle velocities.

\subsubsection{The radial drift barrier}\label{sbreak}

Now we will estimate in which regions of the disk and under which
conditions the solid particles can theoretically overcome the radial
drift barrier.

In section \ref{seceps} we have seen that particle coagulation due to
turbulence in the disk can be described by
$\dot{a}=a\Omega_{\mathrm{k}}\epsilon_0$. We define a particle growth
time scale $\tau_{\mathrm{g}}$ by
\begin{equation}
\tau_{\mathrm{g}}=\gamma\frac{a}{\dot{a}}=\frac{\gamma}{\Omega_{\mathrm{k}}\epsilon_0}\;.
\end{equation}
The parameter $\gamma$ measures how much the solid particle has to
grow to cross the particle size region of fast radial drift, i.e. to
overcome the radial drift barrier. We assume this parameter to have a
certain value determined by the disk model and to be a constant
throughout the disk. The largest radial drift velocity in the disk is
approximately given by $c_{\mathrm{s}}^2/V_{\mathrm{k}}$. We define a
radial drift time scale $\tau_{\mathrm{d}}$ by
\begin{equation}
\tau_{\mathrm{d}}=\frac{r}{c_{\mathrm{s}}^2/V_{\mathrm{k}}}\;.
\end{equation}
The ratio between these two time scales is given by
\begin{equation}\label{tscales}
\frac{\tau_{\mathrm{g}}}{\tau_{\mathrm{d}}}=\frac{\gamma}{\epsilon_0}\left(\frac{H}{r}
\right)^2\;.
\end{equation}
In the last step we made use of Eq.~(\ref{dsh}). Now, the particles
may overcome the radial drift barrier if the ratio
$\tau_{\mathrm{g}}/\tau_{\mathrm{d}}$ is smaller than unity, i.e. if
the growth time scales are smaller than the radial drift time
scales. The parameter $\gamma$ is still indefinite.

To specify this parameter we consider Fig.~\ref{eps} in
Section~\ref{seceps}. These simulation results show for which initial
dust-to-gas ratio $\epsilon_0$ the particles break through the meter
size barrier at a certain radius in the disk. We chose the parameter
$\gamma$ in a way that the condition
$\tau_{\mathrm{d}}>\tau_{\mathrm{g}}$ is in agreement with the results
shown in this figure. This leads to $\gamma\approx 12$. With this
value, the particles should overcome the radial drift barrier if the
inequality
\begin{equation}\label{ceps}
\epsilon_0\gtrsim 12\left(\frac{H}{r} \right)^2
\end{equation} 
holds.

The particles, which break through the radial drift barrier in
Fig.~\ref{eps}, have already drifted inwards. For this reason, the
critical value given by Eq.~\ref{ceps} indicates the initial
dust-to-gas ratio for which the particles most likely break through
the radial drift barrier. The sufficient $\epsilon_0$-value to
overcome the radial drift barrier is presumably even lower than this
value.

\subsubsection{Dust mass loss in the disk}\label{dml}

\begin{figure}
\begin{center}
\includegraphics[scale=0.5]{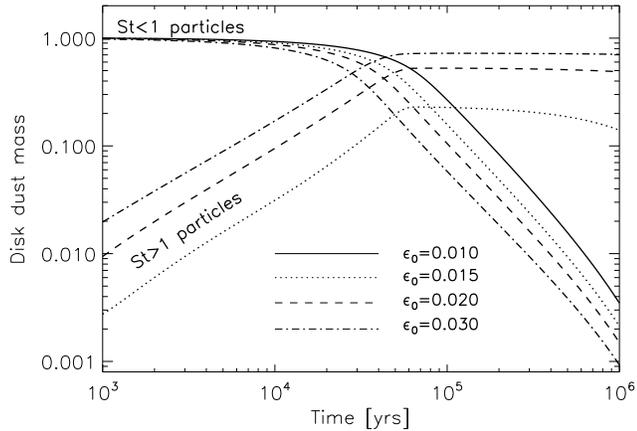}
\caption{The mass of the dust disk for small ($\mathrm{St}<1$) and large
  ($\mathrm{St}>1$) particles between 0.5~AU and 150~AU as a function of time
  for different initial dust-to-gas ratios as discussed in Sec.~\ref{dml}. In
  this simulation only the particle coagulation and the radial motion of the
  dust were considered. Particle fragmentation was neglected.
  \label{massloss}}
\end{center}
\end{figure}

When particles drift into the evaporation zone, they are lost for the
process of planetesimal formation. Hence, the question of how much
solid material is actually lost due to its drift into the inner
regions is of essential importance. We calculate the mass which is
present in small ($\mathrm{St}<1$) and large ($\mathrm{St}>1$)
particles between 0.5~AU and 150~AU as a function of time for
different initial dust-to-gas ratios. The result of this calculation
can be seen in Fig.~\ref{massloss}.

This plot shows, that the mass of the dust disk does not change
significantly within the first $10^4$ yrs for every initial
dust-to-gas ratio considered. Since the power law index of the surface
density is $-0.8$, most of the solid particles are in the outer
regions of the disk. Fast radial drift in the inner disk regions which
takes place in $\sim 10^3$~yrs does not change the total dust mass in
the disk.

After a few $10^4$~yrs, the mass present in small grains starts to
decrease. After 1~Myrs of disk evolution, this mass is less than 1\%
of the initial dust mass. The amount in small grains,
i.e. $\mathrm{St}<1$ particles, is dependent on the initial
dust-to-gas ratio. For $\epsilon_0=0.01$ roughly 0.4\% of the initial
particle mass is present in small grains. For $\epsilon_0=0.03$ this
mass is a factor of 4 lower. Hence, higher initial dust-to-gas ratios
lead to lower dust masses in small grains after 1~Myrs. In the last
section we showed that particles grow faster with increasing
dust-to-gas ratio. Therefore, particles can grow to larger sizes while
moving radially inwards. However, larger sizes also lead to higher
radial drift velocities (Eq.~\ref{pd1}). For this reason, the mass
which is present in small particles in the disk decreases faster for
increasing dust-to-gas ratios. We will find the same behaviour for the
small particles in step 3 where fragmentation is also taken into
account.

For an initial dust-to-gas ratio of 1\%, the mass of the entire dust
disk, i.e. the mass in small and large particles, after 1~Myrs between
0.5~AU and 150~AU is 0.4\% of the initial dust mass. Most of the dust
has drifted into the evaporation zone. For higher initial dust-to-gas
ratios, i.e. higher than 0.015, the particles in the inner regions of
the disk can break through the radial drift barrier. These larger
boulders around 1~AU then sweep up smaller particles which drift
inwards from larger radii (cf. Fig.~\ref{massloss} between $10^3$~yrs
and $\sim 5\times 10^4$~yrs). After $\sim 5\times 10^4$~yrs most of
the dust mass is present in large boulders. While for
$\epsilon_0=0.015$ roughly 20\% of the initial dust mass is present in
$\mathrm{St}>1$ particles after 1~Myrs, the remaining mass in large
boulders is a factor of $\sim 4$ higher for $\epsilon_0=0.03$. Note
that the mass of the remnant dust disk after 1~Myrs changes by a
factor of $\sim 200$ by changing the initial dust-to-gas ratio from
1\% to 3\%. We conclude that the initial dust-to-gas ratio is a
crucial parameter which has an important influence on how much solid
material remains in the disk after 1~Myrs. However, the mass present
in small grains is always less than 0.4\% of the initial dust mass
after 1~Myrs no matter the value of $\epsilon$.

\subsection{Step 3 - Coagulation, radial motion and fragmentation}\label{crf}

We now also include particle fragmentation in our simulation. We 
investigate how this destructive effect influences the particle growth in the
disk and how various disk parameters, like the initial dust-to-gas
ratio, the turbulence parameter $\alpha$ or the fragmentation velocity
$v_{\mathrm{f}}$, influence the coagulation/fragmentation process.

\subsubsection{Time evolution}\label{dde}

\begin{figure}
\begin{center}
\includegraphics[scale=0.73]{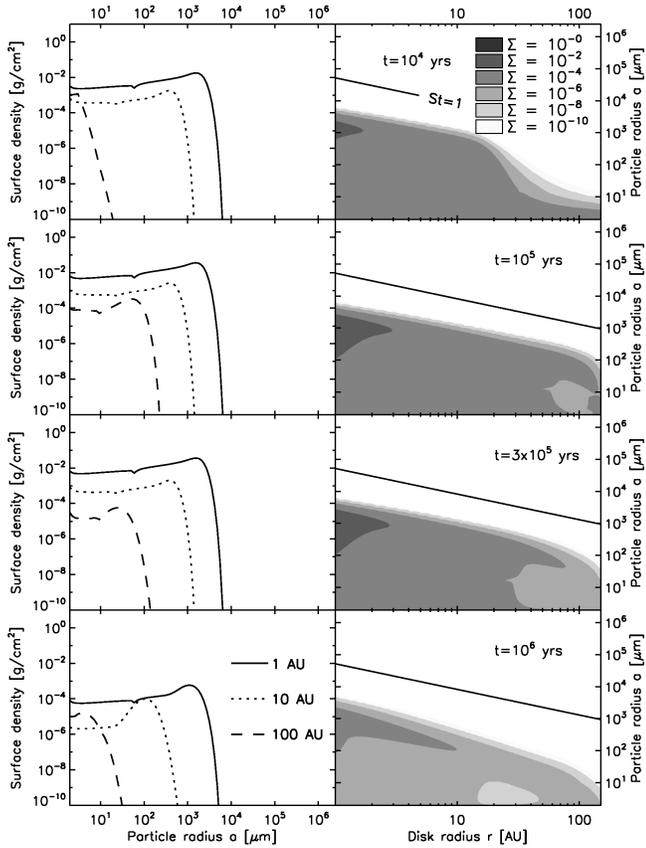}
\caption{As Fig.~\ref{cdrift}, but now also the fragmentation of particles is
  included in the simulations as discusssed in Sec.~\ref{dde}. The left column
  shows the surface density as a function of particle radius at 1, 10 and 100
  AU. The right column shows the corresponding contour plots of the surface
  density as a function of disk radius and particle size.\label{fragplot}}
\end{center}
\end{figure}

The evolution of the disk in the first 1~Myrs is shown in
Fig.~\ref{fragplot}. In this calculation, the fragmentation velocity
is $v_{\mathrm{f}}=10^3$ cm/s and the fragmentation parameter $\xi$ is
1.83. We adopt a disk mass of $10^{-2}$~$M_{\star}$, a turbulent
$\alpha$-value of $10^{-4}$ and an initial dust-to-gas ratio of
$10^{-2}$. The cratering-parameter $\chi$ is 0.5 and $\psi=2$.

After $10^3$~yrs of disk evolution, most of the particles in the disk
$<3$~AU have grown to sizes of some millimeters. However, if
fragmentation is neglected (cf. Sec.~\ref{step2}) the dominant
particle size at 1 AU in the disk after $10^3$~yrs is an order of
magnitude larger. This significant difference is due to the
fragmentation of particles. When the particles reach millimeter size
then destructive effects prevent the particles from growing to larger
sizes (cf. Fig.~\ref{vrel} with 10~m/s). Even after $10^4$~yrs, the
dominant particle size in the disk $<10$~AU is still of the order of a
millimeter. Hence, this particle size corresponds to the fragmentation
barrier for this specific set of disk parameters. Even for long
periods of time the particles are not able to overcome this
barrier. Once the particles have reached the fragmentation barrier the
particle distribution is characterised by an equilibrium between
particle coagulation and particle fragmentation due to destructive
collisions. In other words, the amount of particles of a certain mass,
which are created by dust particle coagulation, equals the amount of
particles, which are destroyed by high velocity collisions. This
steady state will be discussed in more detail later in this Section.

Fig.~\ref{fragplot} indicates that the maximum dominant particle size
$a_{\mathrm{max}}$ and the Stokes number St have the same radial
behaviour. This is due to the fact that relative particle velocities
in our model (except Brownian motion) scale with this dimensionless
number. For this reason, the dominant particle size follows
$a_{\mathrm{max}}\propto r^{-0.8}$ which we obtain directly from the
definition given by Eq.~(\ref{stokesnumber}).

Due to destructive collisions a large amount of dust is present in small
grains as can be clearly seen in Fig.~\ref{fragplot}. We calculate the amount
of dust which is present in grains larger (smaller) than $10^{-2}$~cm after
$10^5$~yrs of disk evolution. While 18\% of the dust mass is present in grains
larger than $10^{-2}$~cm, yet 82\% of the mass is present in smaller
grains. This large population of sub-mm grains should have a strong effect on
the spectrum of the protostellar disk. However, we will not investigate the
influence of the fragmentation parameters, i.e. $v_{\mathrm{f}}$ and $\xi$, on
the disk spectrum which goes beyond the scope of this paper. This will be
investigated in the near future.

\subsubsection{Effect of turbulence}\label{ddz}

\begin{figure}
\begin{center}
\includegraphics[scale=.5]{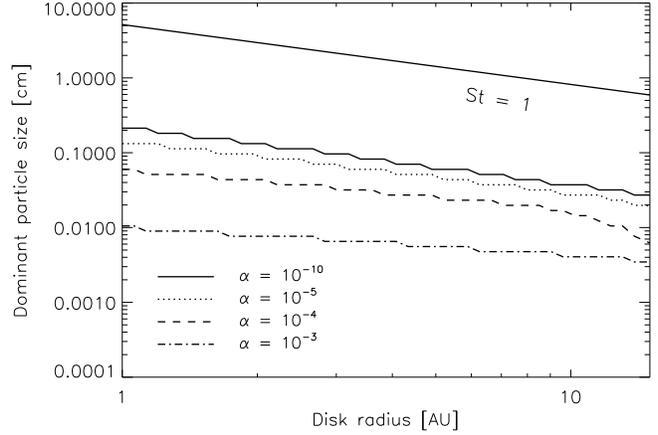}
\caption{The influence of the turbulence parameter $\alpha$ on the dominant
  particle size after $10^4$~yrs of disk evolution for different disk radii
  between 1 and 20~AU as discussed in Sec.~\ref{ddz}. The disk mass is
  $10^{-2}$ $M_{\star}$, the fragmentation velocity is $10^{3}$~cm/s and the
  initial dust-to-gas ratio is $10^{-2}$. This graph also shows the particle
  size for which the Stokes number ist unity. The $\chi$ parameter is set to
  0.5 and $\psi=1$.
  \label{fragvl}}
\end{center}
\end{figure}

Different turbulent $\alpha$-values should lead to different maximum
particle sizes due to destructive collisions. To investigate the influence of
turbulence on the fragmentation barrier, we calculate the dominant
particle size for different $\alpha$-values after $10^4$~yrs of disk
evolution. In this simulation the disk mass is $10^{-2}$~$M_{\star}$, the
fragmentation velocity is $10^{3}$~cm/s, the initial dust-to-gas ratio is
$10^{-2}$ and the results of the calculation are shown in Fig.~\ref{fragvl}.

According to this plot, the dominant particle size is fairly dependent
on $\alpha$ in moderately turbulent disks. If $\alpha$ is changed from
$10^{-3}$ to $10^{-4}$ then the dominant particle size
$a_{\mathrm{dom}}$ changes by a factor of $\sim 5$. We find that less
turbulence shifts the fragmentation barrier towards larger particle
sizes. Hence, in less turbulent disks particles can grow to larger
sizes than in highly turbulent disks.

However, this statement does not hold for extremely low turbulent
disks. In these disks, turbulence is not the main source for relative
velocities and, hence, the fragmentation barrier should not be
dependent on $\alpha$. If $\alpha$ is smaller than $\sim
(c_{\mathrm{s}}/2V_{\mathrm{k}})^2$ (cf. Eqs.~\ref{pd2} and
\ref{relturbvel}) which is $\sim 10^{-4}$ at 1~AU then relative
particle velocities due to radial motion exceed relative dust particle
motions induced by turbulence. To illustrate this independency we
calculate the dominant particle size after $10^4$~yrs for a disk with
a very low $\alpha$-value of $10^{-10}$. The result of this
calculation is also shown in Fig.~\ref{fragvl}. In this nearly laminar
disk, destructive collsions due to relative drift velocities up to
50~m/s prevent particle growth to sizes of more than $\sim 2$~mm at
1~AU.

Relative radial drift velocities are always due to particle size
differences. Monodisperse distributions do not show relative radial
motion. The simulation result for extremely low turbulent disks
($\alpha=10^{-10}$) raises the question how the particle size
dispersion of the dust distribution can produce such high relative
velocities to inhibit particle growth to larger sizes\footnote{We
  define the particle size dispersion as the half-width of the size
  distribution}.
\begin{figure}
\begin{center}
\includegraphics[scale=0.5]{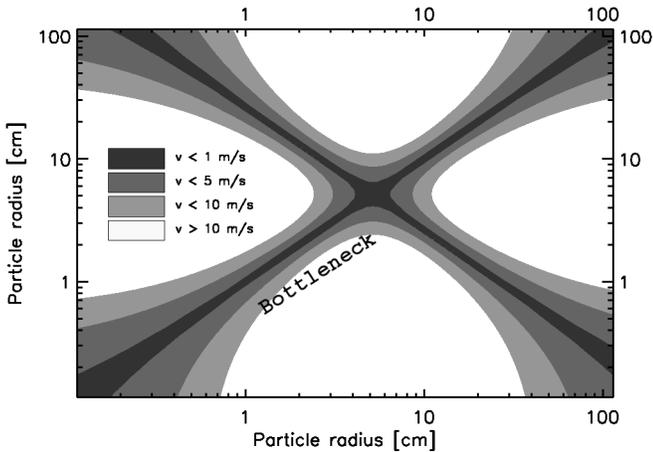}
\caption{The relative particle velocities at 1~AU as a function of
  particle radius as discussed in Sec.~\ref{ddz}. The turbulence
  parameter $\alpha$ in this calculation is $10^{-10}$. This means
  that relative radial motion is the main source for relative particle
  velocities. A critical fragmentation velocity of 10~m/s results in a
  very narrow band in which particle coagulation is still possible. If
  the particle size dispersion is larger than the extent of this
  bottleneck then particle fragmentation starts to play a
  non-negligible role.
\label{bottleneck}}
\end{center}
\end{figure}

We try to answer this question by considering the relative velocities
of dust particles at 1~AU in the disk as a function of particle
radius, c.f. Fig.~\ref{bottleneck}. In this calculation, we adopt an
$\alpha$-value of $10^{-10}$ which means that relative radial motion
is the main source for relative velocities. According to this figure,
particle coagulation is only possible in a very narrow particle size
interval, i.e. in the dark shaded regions of this plot. If the
particle size dispersion is larger than the extent of this
'bottleneck' then particle fragmentation starts to play a
non-negligible role. With Eq.~(\ref{totdv}) for the radial velocities,
we can estimate the importance of fragmentation for a specific
particle size dispersion. We assume a particle size distribution which
has a surface density maximum at $a_0=3$~cm. If the size dispersion is
larger than 1~cm, then particles start to fragment with 100\%
probability. For a particle size dispersion of 0.5~cm and 0.1~cm, the
fragmentation probability decreases to 50\% and 10\%,
respectively. Hence, only for particle size dispersions of some
millimeters, particles might have the chance to overcome the
fragmentation barrier. For larger size dispersions, the fragmentation
probability is far too high to allow the distribution to pass the
bottleneck shown in Fig.~\ref{bottleneck}.

\begin{figure}
\begin{center}
\includegraphics[scale=0.5]{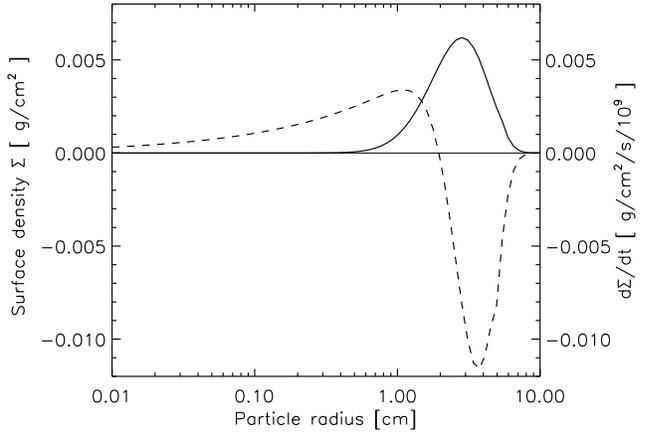}
\caption{The dust particle distribution (solid) and its time
  derivative (dashed) as a function of particle size after 700~yrs of
  evolution at 1~AU in the disk as discussed in Sec.~\ref{ddz}. The
  distribution is located around $a_0=3$~cm and it has a size
  dispersion of $\sim 1.5$~cm. In these 700~yrs, particle
  fragmentation was neglected. In the calculation of the source terms,
  which are shown in this figure, fragmentation is included. These
  source terms show that destructive collsions would rapidly shift the
  dominant particle size to smaller values.
\label{sources}}
\end{center}
\end{figure}

To investigate if the particle size dispersion is narrow enough to
overcome the fragmentation barrier, we consider the following. We
simulate 700~yrs of dust particle evolution neglecting
fragmentation. The result of this simulation, i.e. the particle
distribution at 1~AU in the disk as a function of particle size, is
shown in Fig.~\ref{sources} (solid line). The size dispersion of this
particle distribution is $\sim 1.5$~cm.
Now, particle fragmentation tends to smear out the dust distribution
and it increases the particle size dispersion. Hence, the distribution
shown is Fig.~\ref{sources} represents a best case scenario; the
distribution can not become narrower. What happens if we now switch on
fragmentation? Fig.~\ref{sources} also shows the time derivative of
the particle distribution (dashed line) if fragmentation is
considered. This curve indicates, that destructive collsions would
rapidly shift the dominant particle size towards smaller values.  The
size dispersion is apparently too large for the particles to pass
through the fragmentation bottleneck without undergoing substantial
destructive collsions. If fragmentation is included in the simulations
from the very beginning, then the size dispersion is even larger and,
therefore, the chance of passing the narrow region of coagulation
becomes even smaller.

For a fragmentation velocity of 10~m/s, which we adopt in these
simulations, the particles never overcome the fragmentation barrier,
regardless of the amount of turbulence in the disk since the radial
drift always accounts for destruction. We find that this statement
also holds for larger $\psi$-values. We conclude that the amount of
turbulence in the disk alone does not determine whether particles can
break through the fragmentation barrier or not. Note, that the maximum
radial drift velocity of particles is independent of radius, so that
these statements hold everywhere in the disk.

There are possible scenarios in which the radial particle velocity,
which is the main reason for particle fragmentation in low turbulent
disks, is lower than in the model discussed here. In these cases,
particles might overcome the fragmentation barrier. One possibility
are local gas pressure fluctuations. Since the radial drift velocity
is proportional to the radial gas pressure gradient, local gas maxima
can slow down and even prevent radial particle motion. Therefore, we
expect dust coagulation instead of dust fragmentation in these
maxima. Also local dust particle enhancements can slow down radial
drift. \cite{JohKlaHen06} have shown that the radial drift velocity
can be reduced by a factor of around 2. Further investigations of
particle growth under these conditions, which make particle
fragmentation less likely, are needed.

\subsubsection{Effect of the fragmentation velocity}\label{e_vf}

\begin{figure}
\begin{center}
\includegraphics[scale=.5]{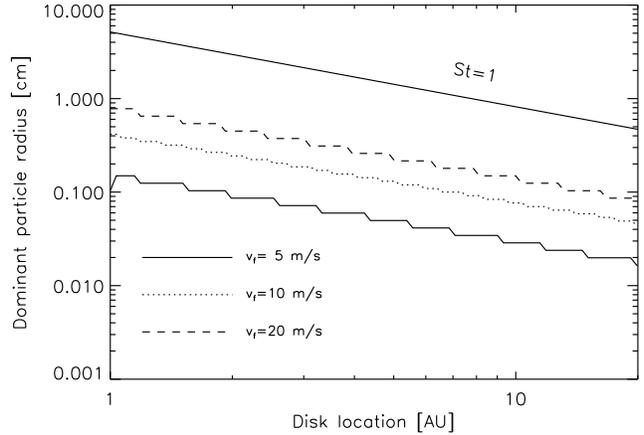}
\caption{The dominant particle size as a function of disk location for 3
different fragmentation velocities after $10^4$~yrs of disk evolution as
discussed in Sec.~\ref{e_vf}. In this simulation, $\psi=2$, $\chi=0$,
$\epsilon_0=0.03$ and the turbulent $\alpha$-value is $10^{-5}$.
\label{frv}}
\end{center}
\end{figure}

For which critical fragmentation velocities can particles break
through the fragmentation barrier? To answer this question let us
consider a best case scenario. We adopt a low turbulent disk, i.e. a
disk in which the relative radial velocities exceed the relative
turbulent particle velocities, and we neglect the effect of cratering
for the moment. We calculate the dominant dust particle size as a
function of disk location for 3 different fragmentation velocities
after $10^4$~yrs of disk evolution. The results of this calculation
can be seen in Fig.~\ref{frv}. In this simulation, the $\alpha$-value
is $10^{-5}$, $\chi=0$ (no cratering) and $\psi=2$.
  
For a fragmentation velocity of 5~m/s, particles can grow to millimeter size
at $\sim 1$~AU in the disk before destructive collisions prevent further
particle growth. In the outer regions, i.e. at 10~AU, the dominant particle
radius is a factor of 10 smaller.  Even for a relatively high critical
velocity of 20~m/s the particles are not able to grow beyond a centimeter at
1~AU.

\begin{figure}
\begin{center}
\includegraphics[scale=.73]{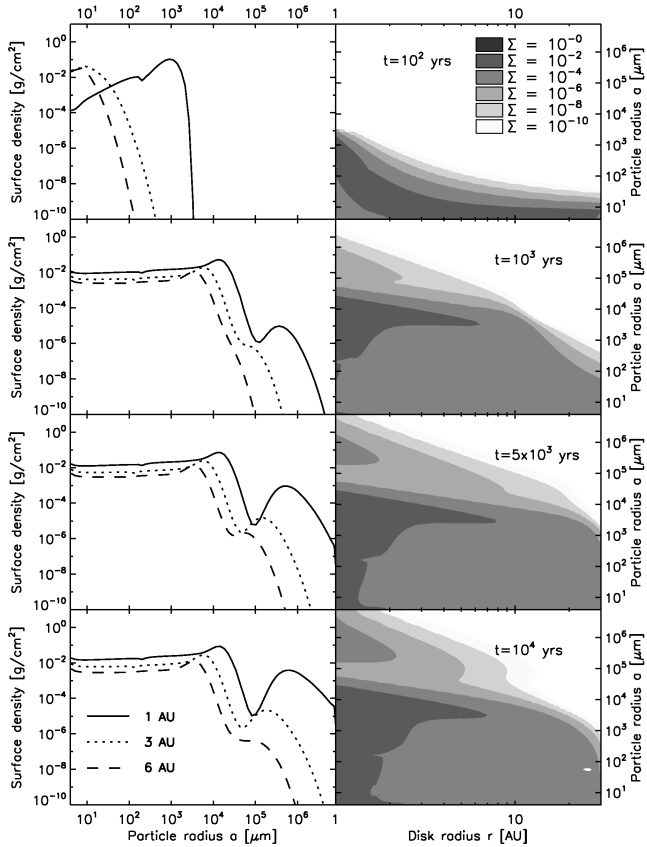}
\caption{These plots show how the particles break through the radial drift
barrier and the fragmentation barrier as discussed in Sec.~\ref{e_vf}. Shown
are contour plots of the surface density as a function of disk radius and
particle radius at 4 different times of disk evolution. The fragmentation
velocity is chosen to have the relatively high value of 30~m/s. In this
simulation $\psi=2$ and $\chi=0$. The initial dust-to-gas ratio is 0.03.
\label{f_break}}
\end{center}
\end{figure}

For even higher fragmentation velocities, i.e. $v_{\mathrm{f}}\sim
30$~m/s, solid particles start to break through the fragmentation
barrier. Fig.~\ref{f_break} shows the dust particle distribution for
this critical velocity as a function of disk radius and particle
radius for 4 different times of disk evolution. This plot indicates
that particles have grown to meter size in the inner parts of the disk
after $10^4$~yrs. However, a fragmentation velocity of several 10~m/s
for centimeter- or even meter-sized boulders is at least
questionable. For lower (and probably also more realistic) critical
velocities, i.e. velocities of $1\ldots 10$~m/s, we never find solid
particles in our simulations which are able to overcome the
fragmentation barrier for any disk parameters considered. For
$\alpha$-parameters, which are higher than the adopted value of
$10^{-5}$ in the simulations of this paragraph, it is even more
unlikely that solid particles may grow to larger sizes. This chance
does not increase if destructive effects due to cratering are also
taken into account.

\subsubsection{Disk dust mass}\label{frag_mass}

\begin{figure}
\begin{center}
\includegraphics[scale=.5]{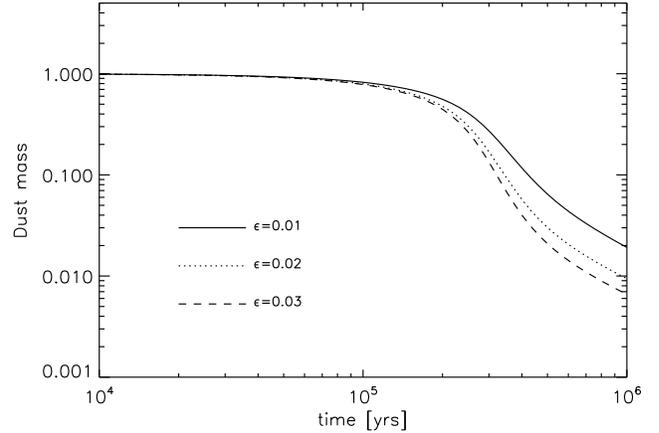}
\caption{The mass of the dust disk between 1~AU and 150~AU as a function of
time for 3 different initial dust-to-gas ratios as discussed in
Sec.~\ref{frag_mass}.  In this simulation, particle growth particle
fragmentation and radial motion are included. The initial disk mass of gas and
solid material is $10^{-2}$~M$_{\star}$, $\alpha=10^{-4}$, $\chi=0.5$,
$\psi=2$ and $v_{\mathrm{f}}=10$~m/s.
  \label{frag_m}}
\end{center}
\end{figure}

As in Sec.~\ref{dml}, we calculate the solid material mass in the disk
as a function of time, but now with the effect of particle
fragmentation included in the simulations. The result of this
calculation is shown in Fig.~\ref{frag_m}.
  
The dust mass does not change significantly within the first
$10^5$~yrs for any $\epsilon_0$ considered. This is the same behaviour
as in the case of no fragmentation. After $10^5$~yrs the mass starts
to decrease rapidly. For an initial dust-to-gas ratio of 0.01 only 2\%
of the initial solid material mass between 1 and 150~AU remains after
1~Myrs. Higher initial dust-to-gas ratios lead to less solid material
after 1~Myrs. For example, for $\epsilon=0.03$ the mass is only 0.7\%
of the initial dust mass, which is a factor of $\sim 3$ lower.

Let us compare the solid material mass after 1~Myrs for
$\epsilon_0=0.01$ in the case of fragmentation/no fragmentation. We
find that the remaining dust mass is a factor of 5 higher if we allow
the particles to destroy each other. This difference is due to
destructive collisions which lead to large amounts of small particles
in the disk (cf. Sec.~\ref{dde}). These small dust grains have low
radial drift velocities and, hence, long radial drift time scales. In
other words, small particles stay much longer in the disk before they
evaporate in the inner regions of the disk. For this reason, the solid
material mass after 1~Myrs is higher in the case of fragmentation than
in the case of no fragmentation.

In the previous sections we found that if fragmentation is included in the
simulations then the dust particles are not able to break through the meter
size barrier. No larger particles in the inner parts of the disk can form
which can sweep up smaller dust particles drifting inward from the outer
regions. For this reason, most of the solid material after 1~Myrs has drifted
into the evaporation zone and is lost for the process of planetesimal
formation.

\subsubsection{Effect of disk model}\label{dmp}

In the introduction we mentioned that the disk model adopted in this
paper differs significantly from the MMSN model. This leads to the
question of how the results of this paper change if different disk
models are considered. In this section, we repeat simulations of
Sec.~\ref{crf} with other disk model parameters, attempting to unveil
the basic changes in the dust particle distribution. Table 1 shows the
disk parameters for the simulations in this section. Model A and B are
the MMSN model and the disk model in this paper, respectively. Model C
is our model, but now with 10\% disk mass instead of 1\% compared to
$M_{\star}$. This leads to gas densities which are comparable to those
of the MMSN model. The mass distribution, however, has a much flatter
radial dependency. The models D to F are the same as A to C, but with
a steeper radial temperature dependency. \cite{andwill07}
observationally find radial temperature profiles with a median power
law index of 0.62. This is slighly higher than the passively
irradiated disk profile of 0.5 adopted in our model.

\begin{table}
\label{diskp}
\begin{tabular}{|c|c|c|c|}
\hline Model & Surface density & Disk & Temperature \\ & power law
index $\delta$ & mass & power law index $\beta$\\ \hline\hline & & &
\\ A & 1.5 & 0.01 & 0.50\\ B & 0.8 & 0.01 & 0.50\\ C & 0.8 & 0.10 &
0.50\\ D & 1.5 & 0.01 & 0.62\\ E & 0.8 & 0.01 & 0.62\\ F & 0.8 & 0.10
& 0.62\\ \hline
\end{tabular}
\caption{Disk parameters for the simulations performed in
  Sec.~\ref{dmp}. The quantity $\beta$ denotes the temperature power
  law index $T\propto r^{-\beta}$. The Models A and B correspond to
  the MMSN model and the model adopted in this paper,
  respectively. Model C is as the model in this paper but now with
  10\% disk mass. The Model D to F are as A to C but with a slighly
  steeper radial temperature dependency.}
\end{table}

Before we come to the results of the simulations, we will
qualitatively discuss the difference between the MMSN model and the
model in the paper at hand. The gas mass densities of our model are
generally smaller than those of the MMSN model. This has the following
main implications. First, solid particles are less coupled to the
motions of the gas. The coupling between the gas and the dust can be
described by the Stokes number St, which is given by $\mathrm{St}=
\rho_{\mathrm{s}}a/\Sigma$. If the surface density of the gas $\Sigma$
decreases, then St is shifted towards higher values. Therefore, the
particle growth barrier due to radial drift and particle
fragmentation, which is usually referred to as the 'meter size
barrier' and which corresponds to the particle radius implied by
$\mathrm{St}=1$, is shifted towards lower particle radii. In the MMSN
model, particles with a Stokes number of unity have radii of $\sim
2$~m at 1~AU in the disk. A surface density slope of $\delta=0.8$
implies $a\sim 5$~cm for $\mathrm{St}=1$ particles at 1~AU. While it
seems challenging to grow particles larger than meter in size in the
MMSN disk model, it is difficult to grow particles larger than
centimeter size in the disk model adopted in the paper at hand.

Second, if the Stokes number is shifted towards higher values then all
quantities depending on this number are influenced by this change as
well. For example, for Stokes numbers smaller than unity the radial
drift velocity of solid particles in the disk is proportional to the
Stokes number, $v_{r}\propto \mathrm{St}$ \citep{Wei77}. Now, if the
Stokes number is modified due to a change of $\delta$ then also the
radial drift of the dust is significantly affected. The Stokes number
also determines relative dust particle velocities in turbulent disks
and, hence, dust particle growth time scales and the maximum dust
particle size due to fragmentation.

\begin{figure}
\begin{center}
\includegraphics[scale=0.73]{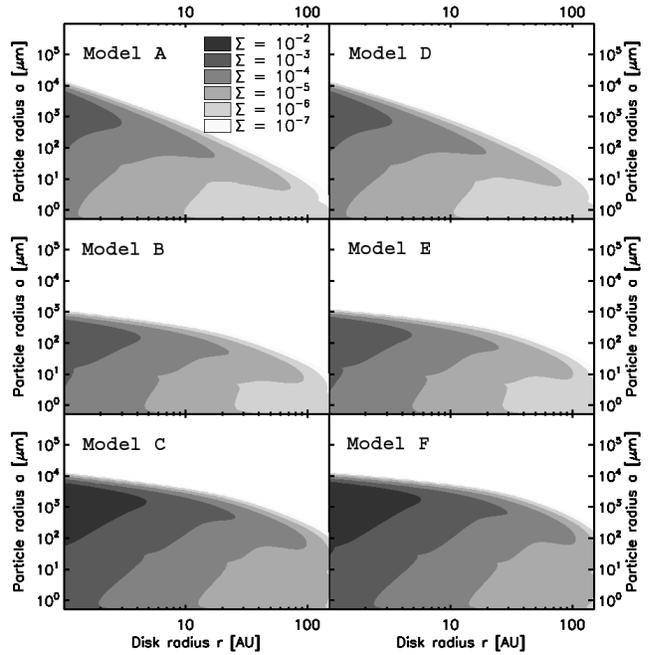}
\caption{The particle distribution in the disk after 1~Myrs for the
  disk models A to F as discussed in Sec.~\ref{dmp}. The Models A and
  B correspond to the MMSN model and the model adopted in this paper,
  respectively. Model C is as the model in this paper but now with
  10\% disk mass. The Model D to F on the right side have a slighly
  steeper radial temperature dependency.
  \label{modelsim}}
\end{center}
\end{figure}

Fig.~\ref{modelsim} shows the particle distribution after 1~Myrs of
disk evolution for the Models A to F. In these simulations, particle
growth, radial particle motion and destructive collisions are
included. The initial dust-to-gas ratio is $10^{-2}$ and the
$\alpha$-value is $10^{-3}$. The $\psi$-parameter is chosen to be 2
and the cratering parameter $\chi=0.5$.
This figure shows that particles can grow to much larger sizes in
model A than in model B in the inner parts of the disk. This is due to
higher gas densities in the MMSN model which alter the Stokes number
and shift the whole particle growth problem towards larger particle
radii. At 1~AU, the gas density in model A is a factor of $\sim 15$
higher than in model B. The dominant particle size before
fragmentation inhibits further particle growth is 3~mm in model A and
0.2~mm in model B. This dominant particle size difference from one
model to the other nicely mirrors the gas density difference between
the two models. Hence, we find that the dominant particle size is
directly proportional to the gas density.

Model C is the same as the model in our paper (B), but now with 10\%
disk mass instead of 1\% compared to $M_{\star}$. Fig.~\ref{modelsim}
shows that the dominant particle radius due to destructive collisions
is shifted by a factor of 10 towards larger particle sizes. According
to these results, particles can grow to a few millimeter in size in
high mass disks before particle fragmentation prevents further
growth. However, even in these very high mass disks, particles can not
overcome the fragmentation barrier. Since the whole
coagulation/fragmentation process scales with gas density, higher disk
masses do not provide a solution for planetesimal formation. The
entire particle growth problem is only shifted towards larger particle
radii.

The right column shows the results of the three simulations A-C if the
radial temperature dependency follows $T\propto r^{-0.62}$
corresponding to the observational median. We do not find a
significant difference in the maximum particle size between these two
model sets.

\begin{figure}
\begin{center}
\includegraphics[scale=0.5]{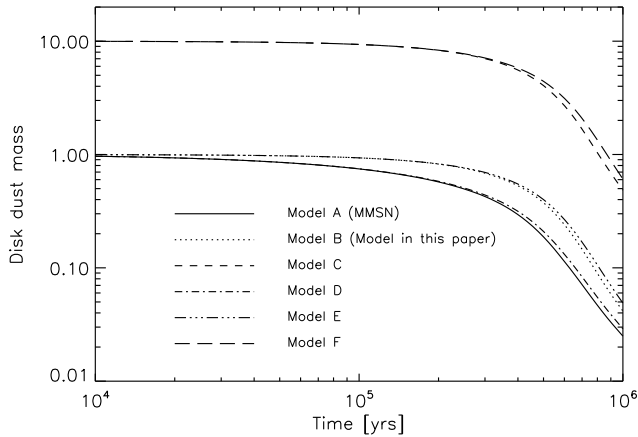}
\caption{The mass of the dust disk as a function of time for the 6
  different disk models A to F as discussed in Sec.~\ref{dmp}. The
  models with 1\% disk mass are normalised to unity. The disk models
  with 10\% disk mass are normalised a factor of 10 larger.
  \label{modmass}}
\end{center}
\end{figure}

We also calculate the mass of the dust disk which is shown in
Fig.~\ref{modmass}.
This plot shows that the remaining dust mass after 1~Myrs of disk
evolution is smaller in the MMSN model than in the model adopted in
this paper. This is due to the fact that the maximum radial drift
velocity is proportional to the power law index $\delta$ of the
surface density profile (cf. Eq.~\ref{pd2}). Since the parameter
$\delta$ is larger in the MMSN model than in our model, the maximum
radial drift speed is also larger. A higher drift speed leads to
shorter drift time scales and, hence, reduces the remaining amount of
dust after a certain time.

In the disk models D-F, the temperature is generally smaller than in
the models A-C. Therefore, the radial drift velocity is also smaller
since $v_{\mathrm{n}}\propto T$. Hence, the disk dust mass in the
model A-C after a certain time is generally smaller than in the models
D-F. Finally, we find that less than 6\% of the initial dust mass is
left after 1~Myrs of disk evolution in any disk model considered.

\subsubsection{Effect of cratering}\label{crater}

If a smaller particle of mass $m_s$ collides with a larger body at a
sufficiently high velocity then the smaller particle does not only fragment
due to this destructive collision but it can also excavate a certain amount of
matter $m_c$ from the larger body, i.e. $m_c=\chi m_s$. This effect is called
cratering. In the following we investigate if this process has an effect on
the equilibrium particle distribution between particle coagulation and
particle fragmentation.

\begin{figure}
\begin{center}
\includegraphics[scale=.49]{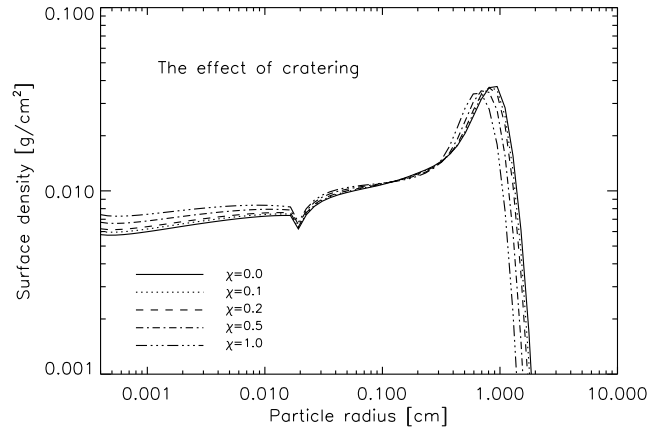}
\caption{The effect of cratering on the equilibrium particle distribution as
discussed in Sec.~\ref{crater}. Shown is the surface density of solid
particles at 1~AU in the disk after $10^4$~yrs of disk evolution for different
cratering-parameters $\chi$.  In this simulation, the fragmentation velocity
is 20~m/s, $\epsilon=0.03$, $\alpha=10^{-5}$ and $\psi=2$.
\label{craterpic}}
\end{center}
\end{figure}

Fig.~\ref{craterpic} shows the equilibrium particle distribution at
1~AU in the disk after $10^4$~yrs of disk evolution for different
cratering-parameters $\chi$.  The fragmentation velocity is 20~m/s,
$\epsilon=0.03$, $\alpha=10^{-5}$ and $\psi=2$.

This plot shows that the equilibrium distribution is hardly affected
whether the effect of cratering is included in the simulations or
not. Changing the $\chi$-value from 0 (no cratering) to 1 (the
projectile particle excavates a crater corresponding to its own mass)
changes the surface density for 50~$\mu$m-sized dust grains by a
factor of 1.3 at most. The maximum peak of the surface density is
shifted from 9~mm to 6~mm by including cratering. We also investigated
the effect of cratering for different fragmentation velocities. In any
case we found that cratering does not significantly affect the
particle distribution. This shows that the main destruction by
fragmentation is due to collisions between particles of not large mass
ratio.

For laboratory experiments which investigate the collision of
particles for astronomical purposes it is interesting to know which
particles collide with which particles in the equilibrium solution
discussed in Sec.~\ref{dde}. For this reason, we calculate collision
rates between dust particles in the disk in Appendix~\ref{sstate} and
we show which collisions are most important for growth/fragmentation.

\section{Discussion and conclusions}

In this paper we study the evolution of the dust in a protoplanetary
disk. We expanded on the work done by \cite{DD05} by adding an
improved treatment of dust aggregate fragmentation and by including
the self-consistent radial drift and mixing of dust aggregates. In
addition to this, our integration method is faster by orders of
magnitude than the original work by Dullemond \& Dominik, by virtue of
two new methods. The first of these is the integration of the
equations in vertical direction without deleting the vertical
structure information, which can be done because the vertical
sedimentation-mixing equilibrium is assumed to be known at all
times. Secondly, we integrate the coagulation-fragmentation in time
using implicit integration, allowing us to overcome the extreme time
scale stiffness that prevented Dullemond \& Dominik from performing
full scale models including fragmentation over time scales of millions
of years.

Using these models we are able to study the problem of the radial
drift and fragmentation barriers in full detail. Our models show that,
consistent with current beliefs, the combination of radial drift and
fragmentation is a strong limitation to growth of aggregates in
disks. Typically, aggregates cannot grow to sizes larger than
millimeters throughout the disk if threshold fragmentation velocities
up to several m/s are considered, though the precise maximum size is
disk model dependent. This statement holds regardless of the amount of
turbulence in the disk. For highly turbulent disks, it is the
turbulence-induced relative velocities that cause the damage to
aggregates, while for nearly non-turbulent disks it is the
differential radial drift that limits the growth. Only if we set the
fragmentation threshold velocity to the unlikely value of 30~m/s or
larger do we find that the fragmentation barrier can be broken and
particles grow to larger sizes. Whether this high fragmentation
threshold is realistic remains to be verified by high-speed laboratory
collision experiments. It has been tentatively shown that high-speed
impacts of a small projectile on a large target may result in growth
\citep{wurm05}, but further study in this direction is imperative.
 
We have also investigated what happens when, for some reason, no
fragmentation occurs. We then find that even in this reduced model, the
particle radius never exceeded several centimeters at any time at any radius
in the disk because of radial particle motion. However, we demonstrated that
this radial drift barrier problem is very sensitive to slight changes in the
initial dust-to-gas ratio. If slighly higher initial dust-to-gas ratios than
the canonical value of 1\% are adopted in the simulations, then particles
can grow to very large sizes in the inner parts of the disk.

We do not include non-linear feedback of the dust back onto the gas in
our model. It has been recently shown by \cite{andersnat07} that such
feedback can lead to the rapid formation of gravitationally bound
clumps of dust which subsequently form Ceres-size bodies. The `dust'
particles, however, must be large (Stokes number near unity) before
this scenario can take place. We find that for low-turbulent disks
Stokes numbers larger than 0.1 can be reached, but we need further
investigation if the amount of dust present in these large grains is
sufficient to trigger such a gravitational collapse in locally
overdense regions in the midplane of the disk.

In the full model, most of the solid material has drifted into the
evaporation zone after 1~Myrs and the remnant disk contains less than
5\% of the initial dust mass. However, there are reasons to believe
that the strong radial drift of particles in such disks may be reduced
by non-linear hydrodynamic effects.  Tentative results from
\cite{JohKlaHen06} find a reduction by a factor of 3 in MRI
turbulence. Moreover, \cite{brauer07} show that observations of
millimeter grains in the outer regions of protoplanetary disks
indicate that the standard radial drift formulae are inconsistent with
the observations. Hence, it may be important to investigate what
happens to our model if radial drift velocities are reduced by some
factor. This will be the topic of a future paper.

The dynamic calculations of \cite{bargesomm95} and \cite{KlaHen97}
also suggest that particle trapping in pressure maxima could be a
solution also to the relative velocity fragmentation barrier. We
intend, in future work, to study how the dust evolves if such pressure
maxima are present.

By being able to model the dust evolution in a self-consistent way,
our code may act as a basis upon wich further modeling of disks is
done.  For example, models of chemistry in disks are very dependent on
the total surface area of dust grains per unit volume, in particular
when grain surface chemistry is taken into account
\citep{aikawa99,sem06}. Models of MRI turbulence in disks depend
strongly on the abundance of free electrons. This abundance is also
very dependent on the total surface area of dust \citep{ilg106} and
our model could -- possibly with an ad-hoc reduction factor for the
drift speed -- provide new insight in this kind of modeling. 

Our disk model involves a constant threshold fragmentation velocity
which can be put into question. Laboratory experiments show that this
threshold velocity is dependent on particle size \citep{bm93,para07}
and we indeed work on that topic to investigate if more realistic
aggregate collision models predict a different disk evolution (Brauer
et al. in prep.).

Finally, current state-of-the-art models of disk structure and their
spectra and images relies on ad-hoc prescriptions of the dust spatial
and size distribution. Models of the kind described here will be
linked to radiative transfer calculations to investigate the effect of
grain evolution on disk structure in the near future.

\section*{Acknowledgements}

We wish to thank Anders Johansen, Andras Zsom, Vernesa Smolcic,
Dimitri Semenov and Konrad Tristram for useful discussions and
comments. We also wish to thank the anonymous referee for useful
criticism that helped us improve the paper.

\bibliographystyle{aa} \bibliography{refsv2.bib}

\onecolumn

\appendix

\section{Coagulation algorithms}\label{alg}
\subsection{Podolak algorithm}

In this section we present an algorithm that was first used be \cite{KovOlu69}
in meteorological science. This algorithm conserves mass and particle number
density. It captivates by its simplicity and it is comparatively easy to
implement into a computer code.

Let us assume a mass grid $m_\mathrm{i}$ and a given spatial number density
${N_{\mathrm{i}}}$ of the particles with mass $m_\mathrm{i}$. Two particles of
mass $m_\mathrm{i}$ and $m_\mathrm{j}$ coagulate with a coagulation rate
$Q_{\mathrm{ij}}$, i.e. the number of coagulation events per time, which is
given by
\begin{equation}
Q_{\mathrm{ij}}=N_{\mathrm{i}}N_{\mathrm{j}}K_{\mathrm{ij}}\;.
\end{equation}
The quantity $K_{\mathrm{ij}}$ denotes the coagulation kernel of the particles
in the mass bins i and j. It is given by the product of the collisional cross
section $\sigma_{\mathrm{ij}}$ between the particles and their relative
velocity $v_{\mathrm{ij}}$. High values of this kernel correspond to high
collision rates which means that the particles are subject to a fast growth
process. Low values of this quantity imply slow growth rates.

Now, an important issue appears if non-linear mass grids are considered. The
resulting mass of the coagulation, i.e. the mass
$m=m_{\mathrm{i}}+m_{\mathrm{j}}$, does not neccessarily match with any of the
mass grid points. This means that in general no mass grid point
$m_{\mathrm{s}}$ can be found which satisfies
$m_{\mathrm{s}}=m_{\mathrm{i}}+m_{\mathrm{j}}$. Therefore, we have to divide
the coagulating mass between the nearest mass grid points in some sensible
way.

We assume that the nearest neighbours are given by
$m_{\mathrm{m}}<m<m_{\mathrm{n}}$. With a linear ansatz we split the
coagulation rate $Q_{\mathrm{ij}}$ into a coagulation rate for the mass
$m_{\mathrm{m}}$ and a coagulation rate for the mass $m_{\mathrm{n}}$,
\begin{equation}\label{cnd}
Q_{\mathrm{m}}=\epsilon\;Q_{\mathrm{ij}}\quad\mbox{and}\quad
Q_{\mathrm{n}}=(1-\epsilon)\;Q_{\mathrm{ij}}\;.
\end{equation}
The number density is a conserved quantity in this algorithm since
$Q_{\mathrm{m}}+Q_{\mathrm{n}}=Q_{\mathrm{ij}}$. Much more important than the
conservation of number density is the conservation of mass. We can enforce
this fundamental conservation principle by setting
\begin{equation}\label{cm}
Q_{\mathrm{m}}m_{\mathrm{m}}+Q_{\mathrm{n}}m_{\mathrm{n}}=Q_{\mathrm{ij}}
(m_{\mathrm{i}}+m_{\mathrm{j}})\;.
\end{equation} 
The last expression defines the value of $\epsilon$ which was a free parameter
till now. Inserting Eqs.~(\ref{cnd}) into Eq.~(\ref{cm}) we find that
$\epsilon$ can be written as
\begin{equation}
\epsilon=\frac{m_{\mathrm{n}}-(m_{\mathrm{i}}+m_{\mathrm{j}})}{m_{\mathrm{n}}
-m_{\mathrm{m}}}\;.
\end{equation}
One may translate these $\epsilon$ parameters for every coagulation process
between particles of species i and j into certain coefficients
$C_{\mathrm{ijk}}$ so that the coagulation equation can be expressed in the
form
\begin{equation}\label{smo}
\dot{N}_{\mathrm{k}}=\frac{1}{2}\sum_{\mathrm{ij}}Q_{\mathrm{ij}}
C_{\mathrm{ijk}}-\sum_{\mathrm{i}}Q_{\mathrm{ik}}\;.
\end{equation}
The quantity $C$ is then given by
\begin{displaymath}
C_{\mathrm{ijk}}=\left\{\begin{array}{ll} \epsilon & \qquad\textrm{if
$m_{\mathrm{k}}$ is the largest mass grid point
$<m_{\mathrm{i}}+m_{\mathrm{j}}$}\\ 1-\epsilon & \qquad\textrm{if
$m_{\mathrm{k}}$ is the smallest mass grid point
$>m_{\mathrm{i}}+m_{\mathrm{j}}$}\\ 0 & \qquad\mathrm{otherwise}
\end{array}\right.
\end{displaymath}
In general, more than 90\% of the elements of the matrix $C$ are
zero. Therefore, a lot of computer calculation time can be saved if only the
non-zero elements in the last expression are summed up.

\subsection{Modified Podolak algorithm}

In simulations for protoplanetary disks, we are interested in particle growth
from sub-micrometer in size, i.e. particles that are part of the interstellar
medium, up to several hundred meters. This means that particle masses ranging
from $10^{-12}$ g to $10^{12}$ g are considered, which corresponds to 24
orders of magnitude in mass.

Now, a major problem appears. Intrinsic computer variables, so-called double
precision variables, have an accuracy only up to 14 digits. Since we are
interested in particle growth by more than 20 orders of magnitude, the
accuracy needed for coagulation simulations exceeds the accuracy provided by
the computer. One solution could be the introduction of quadrupole precision
computer variables but this would slow down the simulation speed
significantly. This accuracy issue leads to certain problems for example the
violation of mass conservation or simulation crashes. In order to perform the
coagulation simulations anyway we have to analytically reformulate the Podolak
algorithm at certain points of the numerical scheme.

For this purpose we first have to introduce a number $c_{e}$ in the following
way. We consider the neighboring mass grid points $m_{\mathrm{k-1}}$ and
$m_{\mathrm{k}}$. The number $c_{e}$ is then defined in a way that the
inequality
\begin{equation}
m_{\mathrm{k-1}}+m_{\mathrm{i}}<m_{\mathrm{k}}
\end{equation}
holds for any i which satisfies the condition
$\mathrm{i}\le\mathrm{k}-c_{e}$. In general, the value of $c_{e}$ is dependent
on the index k. 

Now, we reformulate the algorithm. We formally separate the diagonal elements
of Eq.~(\ref{smo}) from the non-diagonal elements.
\begin{equation}
\dot{N_{\mathrm{k}}}=\frac{1}{2}\sum_{\mathrm{i}=1}^{\mathrm{k}}
N_{\mathrm{i}}^2C_{\mathrm{iik}}K_{\mathrm{ii}}+
\sum_{\mathrm{i}=1}^{\mathrm{k}}
\sum_{\mathrm{j}=1}^{\mathrm{i}-1}N_{\mathrm{i}}N_{\mathrm{j}}
C_{\mathrm{ijk}}K_{\mathrm{ij}}-\sum_{\mathrm{j}=1}^NN_{\mathrm{j}}
N_{\mathrm{k}}K_{\mathrm{jk}}\;.
\end{equation} 
In the next step, we consider the second and third summand of the last
expression. In the second term we separate the case $\mathrm{i}=\mathrm{k}$
which leads to
\begin{equation}
\sum_{\mathrm{i}=1}^{\mathrm{k}-1}\sum_{\mathrm{j}=1}^{\mathrm{i}-1}
N_{\mathrm{i}}N_{\mathrm{j}}C_{\mathrm{ijk}}K_{\mathrm{ij}}+
\sum_{\mathrm{j}=1}^{\mathrm{k}+1-c_e}N_{\mathrm{k}}N_{\mathrm{j}}
C_{\mathrm{kjk}}K_{\mathrm{kj}}-\sum_{\mathrm{j}}^NN_{\mathrm{j}}
N_{\mathrm{k}}K_{\mathrm{jk}}\;.
\end{equation}
This can be rewritten as
\begin{align}
&\quad\;\sum_{\mathrm{i}=1}^{\mathrm{k}-1}\sum_{\mathrm{j}=1}^{\mathrm{i}-1}
N_{\mathrm{i}}N_{\mathrm{j}}C_{\mathrm{ijk}}K_{\mathrm{ij}}+
\sum_{\mathrm{j}=1}^{\mathrm{k}+1-c_e}\left(N_{\mathrm{k}}N_{\mathrm{j}}
C_{\mathrm{kjk}}K_{\mathrm{kj}}-N_{\mathrm{j}}N_{\mathrm{k}}K_{\mathrm{jk}}
\right)-\sum_{\mathrm{j}=\mathrm{k}+2-c_e}^{N}N_{\mathrm{j}}N_{\mathrm{k}}
K_{\mathrm{jk}}\nonumber\\
&=\sum_{\mathrm{i}=1}^{\mathrm{k}-1}\sum_{\mathrm{j}=1}^{\mathrm{i}-1}
N_{\mathrm{i}}N_{\mathrm{j}}C_{\mathrm{ijk}}K_{\mathrm{ij}}
-\sum_{\mathrm{j}=1}^{\mathrm{k}+1-c_e}N_{\mathrm{k}}N_{\mathrm{j}}
K_{\mathrm{kj}}\frac{m_{\mathrm{j}}}{m_{\mathrm{k}+1}-m_{\mathrm{k}}}
-\sum_{\mathrm{j}=\mathrm{k}+2-c_e}^{N}N_{\mathrm{j}}N_{\mathrm{k}}
K_{\mathrm{jk}}\nonumber\\
&=\sum_{\mathrm{i}=1}^{\mathrm{k}-1}\sum_{\mathrm{j}=1}^{\mathrm{i}-1}
N_{\mathrm{i}}N_{\mathrm{j}}C_{\mathrm{ijk}}K_{\mathrm{ij}}+
\sum_{\mathrm{j}=1}^NN_{\mathrm{k}}N_{\mathrm{j}}K_{\mathrm{kj}}
D_{\mathrm{jk}}\;,
\end{align}
where the matrix $D$ is given by
\begin{displaymath}
D_{\mathrm{jk}}=\left\{\begin{array}{ll}
-\frac{m_{\mathrm{j}}}{m_{\mathrm{k}+1}-m_{\mathrm{k}}} & \textrm{if
  $\mathrm{j}\le \mathrm{k}+1-c_e$ and}\\ -1 & \textrm{if
  $\mathrm{j}>\mathrm{k}+1-c_e\;.$}
\end{array}\right.
\end{displaymath}
The new coagulation equation now reads
\begin{equation}
\dot{N_{\mathrm{k}}}=\frac{1}{2}\sum_{\mathrm{i}=1}^{\mathrm{k}}
N_{\mathrm{i}}^2C_{\mathrm{iik}}K_{\mathrm{ii}}+
\sum_{\mathrm{i}=1}^{\mathrm{k}-1}\sum_{\mathrm{j}=1}^{\mathrm{i}-1}
N_{\mathrm{i}}N_{\mathrm{j}}C_{\mathrm{ijk}}K_{\mathrm{ij}}+
\sum_{\mathrm{j}=1}^NN_{\mathrm{k}}N_{\mathrm{j}}K_{\mathrm{kj}}
D_{\mathrm{jk}}\;.
\end{equation}
This was one part of rewriting the algorithm. For the other part we regard the
second term of the last expression, especially the term
$\mathrm{i}=\mathrm{k}-1$. We can rewrite this term as follows,

\begin{align}
&\quad\;\sum_{\mathrm{j}=1}^{\mathrm{k}-2}N_{\mathrm{k}-1}N_\mathrm{j}
C_{\mathrm{k}-1,\mathrm{j},\mathrm{k}}K_{\mathrm{k}-1,\mathrm{j}}\nonumber\\
&= \sum_{\mathrm{j}=1}^{\mathrm{k}-c_e}N_{\mathrm{k}-1}N_{\mathrm{j}}
C_{\mathrm{k}-1,\mathrm{j},\mathrm{k}}K_{\mathrm{k}-1,\mathrm{j}}+
\sum_{\mathrm{k}-c_e+1}^{k-2}N_{\mathrm{k}-1}N_{\mathrm{j}}
C_{\mathrm{k}-1,\mathrm{j},\mathrm{k}}K_{\mathrm{k}-1,\mathrm{j}}\nonumber\\
&=\sum_{\mathrm{j}=1}^{\mathrm{k}-c_e}N_{\mathrm{k}-1}N_{\mathrm{j}}
\frac{m_{\mathrm{j}}}{m_{\mathrm{k}}-m_{\mathrm{k}-1}}
K_{\mathrm{k}-1,\mathrm{j}}+\sum_{\mathrm{k}-c_e+1}^{\mathrm{k}-2}
N_{\mathrm{k}-1}N_{\mathrm{j}}C_{\mathrm{k}-1,\mathrm{j},\mathrm{k}}
K_{\mathrm{k}-1,\mathrm{j}}\nonumber\\
&= \sum_{\mathrm{j}=1}^{\mathrm{k}-2}N_{\mathrm{k}-1}N_{\mathrm{j}}
K_{\mathrm{k}-1,\mathrm{j}}E_{\mathrm{jk}} \nonumber\\
&= \sum_{\mathrm{i}=1}^N\sum_{\mathrm{j}=1}^NN_{\mathrm{i}}N_{\mathrm{j}}
K_{\mathrm{ij}}E_{\mathrm{j},\mathrm{i}+1}\theta\left(\mathrm{k}-\mathrm{j}-
\frac{3}{2}\right)\delta_{\mathrm{i},\mathrm{k}-1}\;.
\end{align}

In this Equation the matrix $E$ is given by
\begin{displaymath}
E_{\mathrm{jk}}=\left\{\begin{array}{ll}
\frac{m_{\mathrm{j}}}{m_{\mathrm{k}}-m_{\mathrm{k}-1}} & \textrm{if
$\mathrm{j}\le \mathrm{k}-c_e$ and}\\
\left[1-\frac{m_{\mathrm{j}}+m_{\mathrm{k}-1}-
m_{\mathrm{k}}}{m_{\mathrm{k}+1}-m_{\mathrm{k}}}\right]\theta
\left(m_{\mathrm{k}+1}-m_{\mathrm{j}}-m_{\mathrm{k}-1}\right)&
\textrm{if $\mathrm{j}>\mathrm{k}-c_e\;.$}
\end{array}\right.
\end{displaymath}
With these two reformulations the coagulation equation can be written in the
form 
\begin{equation}
\dot{N}_{\mathrm{k}}=\sum_{\mathrm{ij}}N_{\mathrm{i}}N_{\mathrm{j}}
K_{\mathrm{ij}}M_{\mathrm{ij}}\;,
\end{equation}
where the final coagulation matrix $M$ is given by
\begin{equation}
M_{\mathrm{ij}}=\frac{1}{2}\delta_{\mathrm{ij}}C_{\mathrm{ijk}}+
C_{\mathrm{ijk}}\Theta\left(\mathrm{k}-\mathrm{i}-\frac{3}{2}\right)
\Theta\left(\mathrm{i}-\mathrm{j}-\frac{1}{2}\right)+\delta_{\mathrm{ik}}
D_{\mathrm{ji}}+\delta_{\mathrm{i},\mathrm{k}-1}E_{\mathrm{j},\mathrm{i}+1}
\Theta\left(\mathrm{k}-\mathrm{j}-\frac{3}{2}\right)\;.\nonumber
\end{equation}
In this expression $\Theta(x)$ denotes the Heaviside distribution, which is
zero for $x<0$ and unity for $x>1$.

\section{Vertical integration}\label{vertint}

Coagulation and fragmentation are local processes. This means that the
equations described in the last section have to be solved at every point in
space. The more space grid points are considered the more time-consuming the
computer simulations become. However, under certain conditions the situation
simplifies. In the following we will describe a scheme that can save a
remarkable amount of computational time. 

We consider the coagulation (fragmentation) equation at a certain space point
$z_{\mathrm{p}}$
\begin{equation}\label{ceq1}
\dot{N}_{\mathrm{k}}(z_{\mathrm{p}})=\sum_{\mathrm{ij}}G_{\mathrm{ijk}}
(z_{\mathrm{p}})N_{\mathrm{i}}(z_{\mathrm{p}})N_{\mathrm{j}}(z_{\mathrm{p}})\;.
\end{equation}
Since we are interested in particle growth in protostellar disks we
can adapt the number densities to this special problem. We assume that
at any given time the vertical particle distribution of any given
particle size is given by a settling-mixing equilibrium distribution
(cf. Eq.~\ref{dsh}). This leads to a density $N_{\mathrm{i}}$ of a
particle of size $a_{\mathrm{i}}$ which depends on the height above
the midplane $z$ as
\begin{equation}\label{vstr}
N_{\mathrm{i}}(z)=\frac{\omega_{\mathrm{i}}}{\sqrt{2\pi}h_{\mathrm{i}}}
\exp{\left[-\frac{1}{2}\left(\frac{z}{h_{\mathrm{i}}}\right)^2\right]}\;.
\end{equation}
In this expression the variable $h_{\mathrm{i}}$ denotes the dust scale height
of the particles with mass $m_{\mathrm{i}}$. The quantity
$\omega_{\mathrm{i}}$ is the surface number density of the particles with that
certain mass. Inserting Eq.~(\ref{vstr}) into Eq.~(\ref{ceq1}) and integrating
over height above the midplane $z$ yields
\begin{equation}
\dot{\omega}_{\mathrm{k}}=\sum_{\mathrm{ij}}\omega_{\mathrm{i}}
\omega_{\mathrm{j}}\sum_{\mathrm{p}}
\frac{G_{\mathrm{ijk}}(z_{\mathrm{p}})}{2\pi
  h_{\mathrm{i}}h_{\mathrm{j}}}
\exp{\left[-\frac{1}{2}\left(\frac{z_{\mathrm{p}}}{h_{\mathrm{i}}}\right)^2\right]}
\exp{\left[-\frac{1}{2}\left(\frac{z_{\mathrm{p}}}{h_{\mathrm{j}}}\right)^2\right]}
\Delta z_{\mathrm{p}}\;.
\end{equation}
If we define
\begin{equation}
\tilde{G}_{\mathrm{ijk}}= \sum_{\mathrm{p}}
\frac{G_{\mathrm{ijk}}(z_{\mathrm{p}})}{2\pi
  h_{\mathrm{i}}h_{\mathrm{j}}}
\exp{\left[-\frac{1}{2}\left(\frac{z_{\mathrm{p}}}{h_{\mathrm{i}}}\right)^2\right]}
\exp{\left[-\frac{1}{2}\left(\frac{z_{\mathrm{p}}}{h_{\mathrm{j}}}\right)^2\right]}
\Delta z_{\mathrm{p}}\;,
\end{equation}
the integrated coagulation equation can be written as
\begin{equation}
\dot{\omega}_{\mathrm{k}}=\sum_{\mathrm{ij}}\omega_{\mathrm{i}}
\omega_{\mathrm{j}}\tilde{G}_{\mathrm{ijk}}\;.
\end{equation}
In this way we have integrated the $z$-dimension out without a single
approximation only with the assumption that the vertical
redistribution goes faster than the coagulation/fragmentation. This
reformulation of the coagulation equation has an obvious advantage,
namely instead of solving the coagulation equation at every point in
$z$ the last expression enables us to solve the equation for every
height above the midplane at the same time. If we assume a vertical
grid with 60 grid points the vertical integration speeds up the
computer simulation routine by a factor of 60.

\section{Implicit differencing}\label{idiff}

If fragmentation is included in the simulations then the limiting time
step for the coagulation/fragmentation process tends to be
small. Fragmentation leads to a permanent amount of small
particles. Small particles, however, are associated with short time
scales. Taking these short time scales into account, the time step of
the numerical simulation can not be chosen to be very large. This
argumentation only holds for explicit numerical solvers. For this
reason we have implemented an implicit solver for the
coagulation/fragmentation equation which we will describe in the
following.

The coagulation/fragmentation equation can be written in the form
\begin{equation}\label{ceq}
\bar{f}=\bar{F}(\bar{f})\;,
\end{equation}
where $\bar{f}$ denotes the particle distribution vector on the mass grid and
the function $\bar{F}$ describes the time evolution. In one time step $\Delta
t$ at a certain time $t$, we now want to calculate the new particle
distribution $\bar{f}_n=\bar{f}(t+\Delta t)$ from the old distribution
$\bar{f}_o=\bar{f}(t)$. Therefore we rewrite Eq.~(\ref{ceq}) as
\begin{equation}\label{cc2}
\bar{\epsilon}=\Delta t \bar{F}(\bar{f}_i),
\end{equation}
where $\bar{\epsilon}=\bar{f}_n-\bar{f}_o$ and
$\bar{f}_i=\xi\bar{f}_o+(1-\xi)\bar{f}_n$. The time evolution of the function
$\bar{f}$ with $\xi=1$ is called explicit, while the time evolution with
$\xi=0$ is usually called implicit. Choosing $\xi=0$ in our case, we can
perform a Taylor expansion of the right-hand side of Eq.~(\ref{cc2}) which
leads to
\begin{equation}\label{gha}
\bar{\epsilon}=\Delta t \bar{F}(\bar{f}_o)+\Delta t \tilde{J}\bar{\epsilon}\;.
\end{equation}
The Matrix $\tilde{J}$ denotes the Jacobi matrix which is definded as
$\tilde{J}_{ij}=\partial F_i/\partial f_j$. Solving Eq.~(\ref{gha}) for
$\bar{\epsilon}$ leads to
\begin{equation}
\bar{\epsilon}=\left[1-\Delta t \tilde{J}\right]^{-1}\Delta t
\bar{F}(\bar{f}_o)\;.
\end{equation}
Hence, the evolution of the implicit time step reduces to a matrix inversion
which can be done easily.

\twocolumn

\section{Tests}

This part of the appendix considers a comparison between the model
described in Section~\ref{moddel} and a model which involves the
following approximation. The particle distributions which can be seen
for example in Fig.~\ref{coag} are fairly narrow peaks. Hence, it is
suggestive to approximate the particle distribution by a monodisperse
distribution, i.e. a distribution with only one single particle
size. In this case the coagulation equation simplifies enormously. We
will compare these two different models by considering coagulation due
to Brownian motion and turbulent coagulation. Also the routine for
radial motion will be checked against this simplified model.

\subsection{Simplified model and numerical setup} 

We assume that a certain amount of equally-sized dust particles with radius
$a$ is located at a single radius $r$ in the disk. The surface density of
these particle is given by $\Sigma_{\mathrm{d}}$. The dust particles are
vertically distributed according to Eq.~(\ref{dsh}). For the test cases we
ignore the radial motion of the gas and the radial turbulent diffusion of the
dust so that the radial motion of the dust is given by Eq.~(\ref{pd1}). The
assumption of a monodisperse distribution leads to a coagulation equation
given by Eq.~(\ref{kornetf}) \citep{kornet01}. In the test cases we will use a
surface density of the dust $\Sigma_{\mathrm{d}}=5.26\times 10^{-2}$ g/cm$^2$
and an initial location in the disk given by $r_0=5.53$~AU. Every other
parameter was already mentioned in Section~\ref{moddel}.

\subsection{Radial drift}

\begin{figure}
\begin{center}
\includegraphics[scale=.5]{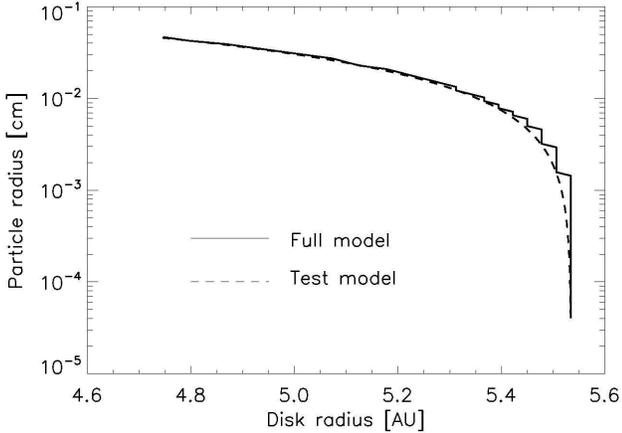}
\caption{Test for the radial drift routine. Coagulation is neglected in this
  simulation. The lines denote the particle position in the disk after
  $10^3$~yrs of radial drift from the initial radial position $r_0=5.53$~AU
  for different dust particle sizes.
  \label{dt}}
\end{center}
\end{figure} 

First, we check the radial drift routine. Any coagulation is neglected. With
the values mentioned above we let the dust particles drift for $10^3$~yrs and
plot their radial position in the disk for different particle sizes.  The
results of this simulation are shown in Fig.~\ref{dt}. In this figure the
solid line correponds to the dominant particle size of the dust distribution
in the full model. The dashed line denotes the result of the monodisperse
model. Any discontinuous effects are due to the radial grid. 

\subsection{Coagulation}\label{monobull}

\begin{figure}
\begin{center}
\includegraphics[scale=.5]{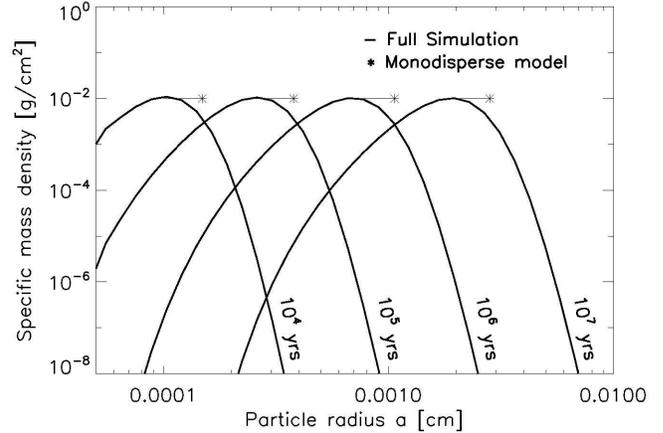}
\caption{Test for the coagulation routine. This plot shows the particle
  distribution in the full model and the monodisperse model for coagulation
  due to Brownian motion at 4 different times. The stars '$\star$' denote the
  particle size in the monodisperse model. The largest discrepancy in particle
  size between the two models is a factor of 1.6 in particle size $a$.
  \label{bt}}
\end{center}
\end{figure}

Now, we consider the coagulation of the dust while the radial motion
of the dust is neglected. We investigate the dust particle coagulation
at $r_0=5.53$~AU in the disk and we first focus on coagulation due to
Brownian motion. The results of both models are shown in
Fig.~\ref{bt}. The stars '$\star$' denote the particle size in the
monodisperse model. The largest discrepancy in particle size between
the two models is a factor of $\sim 1.6$ after $10^7$ yrs.

\begin{figure}
\begin{center}
\includegraphics[scale=.5]{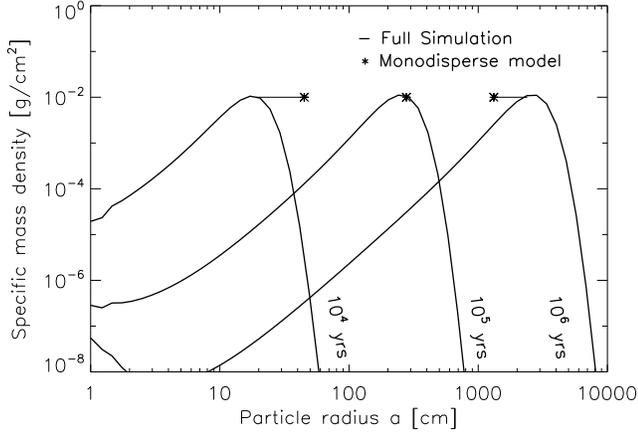}
\caption{Test for the coagulation routine. This plot shows the particle
  distribution in the full model and the monodisperse model for coagulation
  due to Brownian motion and turbulence in the disk at 3 different times. The
  stars '$\star$' denote the particle size in the monodisperse model. The
  largest discrepancy in particle size between the results that are shown
  in this figure is a factor of $\sim 2.7$.
  \label{ct}}
\end{center}
\end{figure}

The results of the same simulation but now with particle growth due to
turbulent coagulation included are shown in Fig.~\ref{ct}. The largest
discrepancy in this case in particle size between the two models is a factor
of $\sim 2.7$ in radius for $t=10^4$~yrs.

However, if the Stokes number of the dust particles is smaller than
unity, then the particle size of both models after a certain time can
differ in more than one order of magnitude. This is due to the
following reason. If the Stokes number is smaller than unity then the
relative turbulent velocity of the dust follows $\Delta v\propto
\sqrt{\mathrm{St}}$. With the coagulation equation~(\ref{kornetf})
this leads to $\dot{a}\propto a$ and its solution $a\propto
\exp(t)$. In the particular case of turbulent coagulation and
$\mathrm{St}<1$, the particle size as a function of time is not given
by a polynomial expression, i.e. an expression of the form $a\propto
t^{\gamma}$, but an exponential function. If a time evolution follows
a polynomial law then two different but similar initial conditions
will lead to different but similar particle sizes after any time. Let
us consider the time evolution
\begin{equation}
a(t)=\left[ct+a_0^{1/{\gamma}}\right]^{\gamma}.
\end{equation} 
First, the initial particle size $a_0$ gets unimportant for large
$t$. Second, two different c-parameters, i.e. $c_1$ and $c_2$, will
lead to particle sizes which will differ by a factor of
$(c_1/c_2)^{\gamma}$. This factor is constant and does not
increase. 

On the other hand, if the particles grow exponentially,
\begin{equation}
a(t)=a_0\exp(ct),
\end{equation}
then the initial particle size $a_0$ will always play a
role. Moreover, two different c-values, i.e. $c_1$ and $c_2$, will
lead to particle sizes which will differ by a factor of
$\exp(c_1t-c_2t)$. This factor increases in time. For example, in the
case of turbulent coagulation (cf. eq.~\ref{solu}) the parameter $c$
is given by $\epsilon_0\Omega_{\mathrm{k}}$. An initial dust-to-gas
ratio of 1\% leads to $a=a_0e\approx 2.72a_0$ after 100 orbits. An
initial dust-to-gas ratio of 2\% after 200 orbits already implies
$a=a_0e^4\approx 54.6a_0$ which is a factor of 20 larger. In the case
of turbulent coagulation and St smaller than unity, small changes in
the intial conditions lead to large differences in the growth
behaviour. Hence, a change from a monodisperse model to a model with a
whole particle dispersion presumably leads to similar effects.

\section{Collision rates}\label{sstate}

In this part of the Appendix, we investigate the equilibrium particle
distribution between coagulation and fragmentation after $10^3$ yrs of
disk evolution at 1 AU in the disk. In this calculation the
fragmentation velocity is $v_{\mathrm{f}}=10^3$~cm/s and the
fragmentation parameter $\xi$ is 1.83. We adopt a disk mass of
$10^{-2}$~$M_{\star}$, a turbulent $\alpha$-value of $10^{-3}$ and an
initial dust-to-gas ratio of $10^{-2}$.  The cratering parameter
$\chi=0.5$ and we adopt $\psi=2$. We focus on the collision rates
between particles of different sizes and how important these
collisions are for particle growth.

\begin{figure}
\begin{center}
\includegraphics[scale=.5]{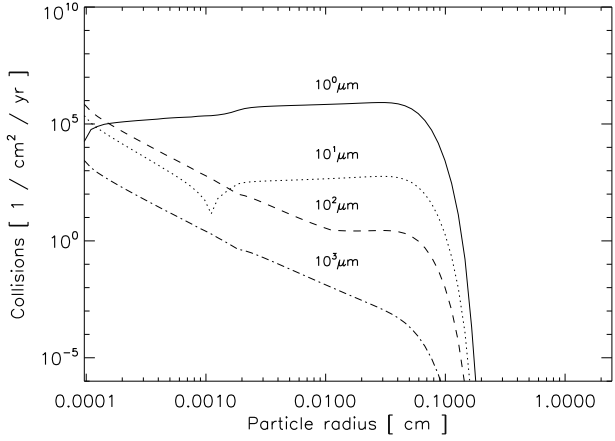}
\caption{This plot shows the vertically integrated number of collisions per
  time at 1~AU in the disk as discussed in Sec.~\ref{sstate}. This calculation
  is based on the particle distribution after $10^3$~yrs of disk evolution
  shown in Fig.~\ref{fragplot}. In this simulation we adopted a disk mass of
  $10^{-2}$ $M_{\star}$ and a turbulent $\alpha$ parameter of $10^{-3}$.
  \label{papprates}}
\end{center}
\end{figure}

We define the collision rates $R(r_1,r_2)$ as the number of collisions per
second between particles of radius $r_1$ and $r_2$ in a vertical column with a
cross section of 1 cm$^2$. These collision rates are shown in
Fig.~\ref{papprates}. The collision rates for particles of equal size decrease
dramatically with increasing particle radius. The collision rates of
micrometer-sized particles and the collision rates of mm-sized particles
differ by more than 10 orders of magnitude. While the collision rate for
1~$\mu$m particles is $\sim 10^{4\ldots 5}$~cm$^{-2}$yrs$^{-1}$, the collision
rate for 10~$\mu$m particles is already more than two orders of magnitude
lower. The smallest particles have the highest collision rates. Since the
coagulation probability for $\mu$m particles is rather high, this leads to
relatively short coagulation time scales for small particles. The collision
rate is $1$~cm$^{-2}$yrs$^{-1}$ for $10^{2}$~$\mu$m particles. For particles,
which are one order of magnitude in radius larger, this rate has already
dropped to $10^{-6}$~cm$^{-2}$yrs$^{-1}$. However, the behaviour of the
collision rates between non-equal sizes particles is different. For example,
the collision rate between $\sim\mu$m-sized dust particles and larger
particles does hardly change over a wide range.

\begin{figure}
\begin{center}
\includegraphics[scale=.5]{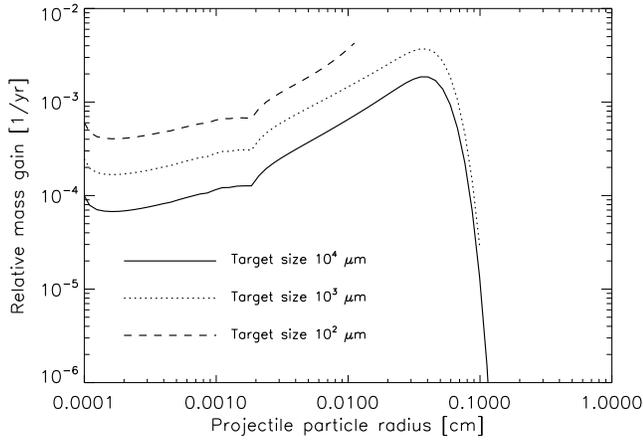}
\caption{This plot shows the relative mass gain per year at 1 AU in the
  disk. This calculation is based on the particle distribution after
  $10^3$~yrs of disk evolution.  The disk mass is $10^{-2}$ $M_{\star}$ and
  the turbulent $\alpha$ parameter is $10^{-3}$.
  \label{paprates}}
\end{center}
\end{figure}

The collision rates, which are shown in Fig.~\ref{papprates}, do not provide
information about the importance of collisions for particle growth. Not only
the number of collisions per time is important for particle coagulation, but
also the mass of the particles itself. For this reason, we calculate the
relative mass gain of a single particle with radius $r_1$. If the number of
all particles of radius $r_1$ in a vertical column is given by $N_1$, then in
average every single particle of size $r_1$ collides with $R(r_1,r_2)/N_1$
particles of size $r_2$ per second. We assume that the sticking probability is
unity. This means that the mass, which the particle of radius $r_1$ sweeps up
in time, is given by $R(r_1,r_2)m_2/N_1$. Its relative mass gain rate is then
given by
\begin{equation}
\Xi_{r_1}(r_2)=\frac{R(r_1,r_2)}{N_1}\frac{m_2}{m_1}\;.
\end{equation}
This quantity is shown in Fig.~\ref{paprates}. Since larger particles sweep up
smaller particles the mass gain rates $\Xi$ are not shown for particle sizes
$r_2>r_1$.

We find that the collisions which are most important for particle
growth are collisions between equal-sized particles. This statement
holds for particles smaller than $\sim 0.5$~mm in size. If the
particle with radius $r_1$ is larger than this value then collisions
with $r_2\sim 0.5$~mm sized particles are most important for the
growth of the dust.


\begin{table*}[!p]
\begin{center}
\begin{tabular}{lr}
Variable & Explanation\\ \hline \hline $a$ & radius of the
particle\\ $m$ & mass of the particle\\ $\rho_{\mathrm{s}}$ & solid
material density of the particle\\ \hline $r$ & distance to the
central star from a point in the midplane\\ $z$ & height above the
midplane\\ T & temperature\\ $c_{\mathrm{s}}$ & isothermal
soundspeed\\ $\Omega_{\mathrm{k}}$, $V_{\mathrm{k}}$ & Kepler
frequency, Kepler velocity\\ $H=c_{\mathrm{s}}/\Omega_{\mathrm{k}}$ &
gas scale height\\ $h$ & dust scale height\\ $\Sigma_{\mathrm{g}}$,
$\Sigma_{\mathrm{d}}$&surface density of the gas and the
dust\\ $\rho_{\mathrm{g}}$, $\rho_{\mathrm{d}}$ & gas and dust
density\\ $\epsilon_0$, $\epsilon$& initial and current dust-to-gas
ratio\\ $r_{\mathrm{in}}$& inner radius of the
disk\\ $r_{\mathrm{out}}$&outer radius of the disk\\ $M_{\star}$&mass
of the central star\\ $M_{\mathrm{disk}}$&mass of the
disk\\ $v_{\mathrm{n}}$&maximum radial drift velocity\\ \hline
$\alpha$, $q$&turbulence parameters\\ $\mbox{St}$& Stokes number of
the particle\\ $D_{\mathrm{g}}$, $D_{\mathrm{d}}$& diffusion
coefficients of gas and dust\\ \hline $v_{\mathrm{dust}}$ & radial
dust drift velocity due to gas drag\\ $v_{\mathrm{gas}}$ & radial gas
accretion velocity\\ $v_{\mathrm{dust}}^{\mathrm{tot}}$ & total radial
dust velocity\\ \hline $\xi$ & slope of the particle distribution
after fragmentation\\ $\chi$ & relative amout of mass removed from the
target particle by cratering\\ $v_{\mathrm{f}}$ & threshold
fragmentation velocity \\ \hline
\end{tabular}
\caption{Important variables used in the course of this paper.}
\end{center}
\end{table*}

\end{document}